# How precisely are solute clusters in RPV steels characterized by atom probe experiments?

N. Castin[(1,*)], P. Klupś[(2)], M. Konstantinovic[(1)], G. Bonny[(1)], M.I. Pascuet[(3)], M. Moody[(2)], L. Malerba[(4)].

[(1)] Studiecentrum voor Kernenergie – Centre D'Études de L'énergie Nucléaire (SCK CEN), Nuclear Energy Technology (NET) department, Boeretang 200, Mol, B2400, Belgium.
[(2)] Department of Materials, University of Oxford, Parks Road, Oxford, OX1 3PH, UK.
[(3)] Gerencia Materiales - CAC, CNEA/CONICET, Godoy Cruz 2290, C1425FQB, CABA, Argentina.
[(4)] Centro de Investigaciones Energéticas, Medioambientales y Tecnológicas (CIEMAT), Avda. Complutense 40, Madrid, 28040, Spain.
[(*)] Corresponding author: ncastin@sckcen.be; nicolas.m.b.castin@gmail.com

## Abstract

Atom probe tomography (APT) is a powerful microscopy technique to characterize nano-sized clusters of the alloying elements in the bulk of reactor pressure vessel (RPV) steels. These clusters are known to dominantly determine the evolution of mechanical properties under irradiation. The results are conventionally summarized as the overall number density $N$ and the average diameter $D$ of the solute clusters identified in the material. Here, we demonstrate that these descriptors are intrinsically imprecise because they are steered by the parameters involved in the measurement and data processing, some of which are directly under the control of the operators, but some others not. Consequently, a direct comparison between data derived at different laboratories is compromised, and key trends such as the evolution with dose, are masked. This study relies on a state-of-the-art physical model for neutron irradiation in steels to make reliable estimates of the true microstructure before the measurement is performed, which allows the prediction of the population of solute clusters that are not seen by APT. We mimic APT measurements from simulated microstructures, performing a detailed study of the effects of the parameters of the analysis. We show that the values of $N$ and $D$ reported in the scientific literature can be matched by the predictions of our theoretical model only if specific sets of parameters are used for each laboratory that issued the measurements. We also show that if, on the contrary, all studied cases are analyzed in a consistent way, the scatter of $N$ and $D$ values is reduced. Specifically, we find that the average diameter $D$ is nearly a constant value with dose, independently of the material's chemical compositions, while $N$ increases with dose, but is also influenced by other variables.

# 1 Introduction

Radiation-induced hardening and embrittlement of reactor pressure vessel (RPV) steels during operation are known to be mainly induced by the segregation of minor alloying elements [1-4]. Atom probe tomography (APT) is a powerful technique to characterize the distribution of atomic species at the nanoscale [6,7], revealing features with a wealth of details that would be challenging to obtain with other techniques [8-12]. This level of detail is instrumental with a view to understanding the physical processes that drive the microstructural evolution of materials in service conditions and are of great help for the validation and calibration of physical models for neutron irradiation in steels [13-23]. Many APT studies for RPV materials revealed that some alloying elements, mainly Ni, Mn, Si, P and Cu, form diffuse, nanosized clusters (average diameter $D \approx$ 2-4 nm) which are usually homogeneously dispersed in bulk (number density $N \approx 10^{22}$-$10^{24}$ m$^{-3}$). These clusters hinder the flow of dislocations under the application of external stresses, thereby affecting the macroscopic properties of irradiated materials. Dispersed barrier hardening models correlate the obstacle strength $\sqrt{ND}$ with the yield strength increase [24], or with the increase of the ductile-to-brittle transition temperature [5]. Consequently, the number density $N$ and the average diameter $D$ of solute clusters in bulk are key indicators for the degradation of nuclear materials under service conditions.

An APT experiment consists of evaporating tens-to-hundreds of millions of atoms from the surface of fine needle-shaped samples; these reach a sensor that can derive their initial position; a mass spectrometer then determines their chemical nature. In this way, the original atomic species distribution is reconstructed, and its analysis enables the identification of solute clusters down to a handful of atoms. The number density $N$ and the average diameter $D$ of these solute clusters can then be derived and are typically reported in scientific manuscripts as main descriptors of the observed microstructure. In contrast, the detailed size distribution of the detected solute clusters is generally not explicitly given.

The reported values of $N$ and $D$ are known to be affected by significant uncertainties [25,26], both stemming from the APT experiment itself and from the method of data analysis. For example, due to the limited single ion detection efficiency in modern instruments, approximately 20%-65% of the atoms in the field-of-view are lost to the analysis. Furthermore, there are limits to spatial resolution [27] and the reconstructed lattice is anisotropically distorted. As a result, the cluster/matrix interface in the APT data is blurred, meaning that it is not straightforward to define the boundaries of the clusters and extract them for further statistical analysis. This introduces subjectivity into the process and, although best practices exist, no standardized approach has been agreed upon by the community, leading to discrepancies between laboratories [28,29].

In this work, we collected a database of eleven sets of APT data from the scientific literature [5,11,30-46], where evidence for such discrepancies is described in what follows. The database is summarized in Tab. 1, and the reported values of N and D are plotted versus the received neutron dose in Fig. 1. Most of the data originate from actual nuclear power plants, but some sets include model alloys and chemically tailored steels. Tab. 1 and Fig. 1 therefore covers a wide range of materials, although all belong to the low-Cu (< 0.1 at%) category, i.e., where no thermodynamically driven precipitation is expected in binary alloys.

**Table 1** – Sets of APT data taken from the literature. For the diameter D, parentheses denote the average values within the set, ±two times the standard deviation. Detailed information on the 74 individual measurements is given in Tab S1-S3 in the supplementary information.

| Name | Number of materials / measurements | Ni content (at%) | Mn content (at%) | Dose (dpa) | APT Number density N ($10^{22}$m$^{-3}$) | APT Average diameter D (nm) | Ref for APT data |
|------|------|------|------|------|------|------|------|
| **Almirall** | 9 / 9 | 0.19 – 3.5 | 0.06 – 1.34 | 0.2 | 9 – 220 | 2.15 – 2.6 (2.36 ± 0.32) | [30] |
| **LONGLIFE** | 4 / 6 | 0.38 – 1.61 | 0.86 – 1.14 | 0.03 – 0.1 | 8.9 – 164 | 2.3 – 3.20 (2.65 ± 0.73) | [5] |
| **Huang, Auger** | 2 / 8 | 0.53 ; 0.68 | 1.26 ; 1.36 | 0.03 – 0.28 | 1.3 – 110 | 3.0 – 4.0 (3.13 ± 0.65) | [31,32] |
| **Jenkins** | 4 / 4 | 1.54 – 3.22 | 0.04 – 1.93 | 0.17 | 70 – 210 | 1.78 – 2.44 (2.07 ± 0.50) | [33] |
| **Meslin, Lambrechts** | 5 / 11 | 0.57 – 0.71 | 0.98 – 1.32 | 0.03 – 0.2 | 10 – 88 | 1.18 – 2.20 (1.49 ± 0.50) | [34-37] |
| **Edmondson, Wells** | 6 / 6 | 0.71 ; 1.68 | 0.4 ; 1.5 | 0.03 – 0.2 | 2 – 53 | 1.00 – 1.80 (*) (1.46 ± 0.51) | [11, 38,39] |
| **Courilleau (VVER)** (**) | 2 / 12 | 1.23 ; 1.68 | 0.35 ; 0.63 | 0.03 – 0.28 | 20 - 409 | 1.8 – 2.4 (2.09 ± 0.37) | [40] |
| **Miller (VVER)** (**) | 2 / 7 | 1.19 ; 1.69 | 0.39 ; 0.69 | 0.04 – 0.26 | 30 – 260 | 1.54 – 1.86 (1.76 ± 0.21) | [41] |
| **Kuleshova (VVER)** (**) | 2 / 3 | 1.18 – 1.6 | 0.47 – 0.75 | 0.08 | 14 – 61 | 2.4 – 2.8 (2.57 ± 0.34) | [42-44] |
| **Dohi** | 1 / 5 | 0.59 | 1.39 | 0.05 – 0.21 | 20 – 33 | 2.2 – 3.10 (2.84 ± 0.84) | [45] |
| **Takeuchi** | 1 / 3 | 0.59 | 1.39 | 0.03 – 0.17 | 4.1 - 23 | 2.14 – 2.82 (2.44 ± 0.70) | [46] |

(*) The D in the Edmondson-Wells set is reported excluding the outlier (D = 3.4nm) from Ref. [38].

(**) VVER is an acronym that denotes RPV materials from Russian-built nuclear reactors (water-water energy reactors). They are similar to Western RPV materials, but they contain some amount of Cr (~3%) and thus have a slightly different microstructure.

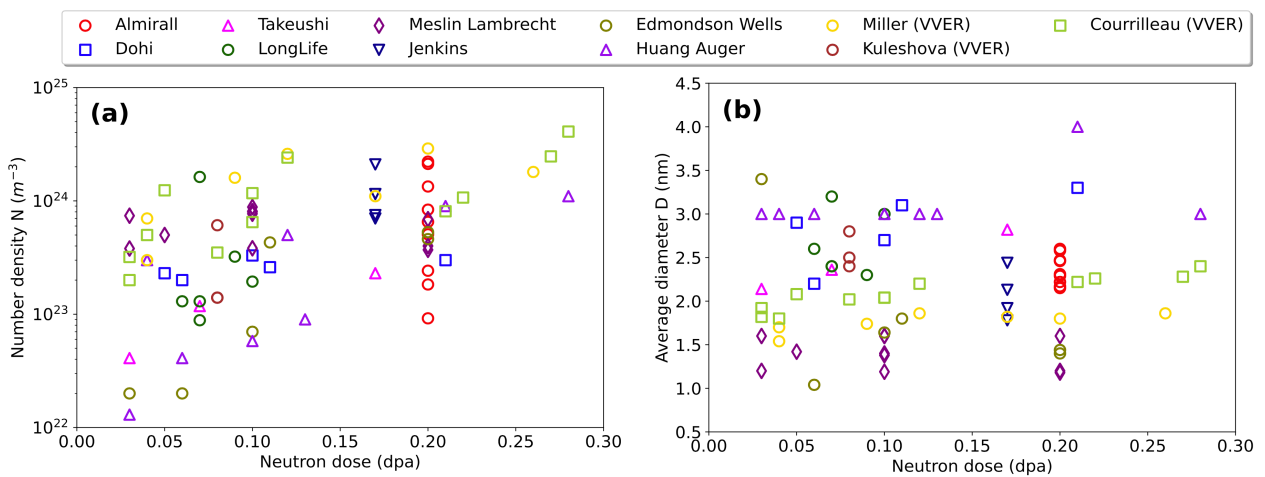

**Fig. 1** – Evolution with neutron dose of (a) the number density, $N$, and (b) the average diameter $D$ of the solute clusters, according to the APT experimental data from the studies listed in Tab. 1. Additional figures are provided as supplementary information and illustrate that no clear correlation can be found between $N$ or $D$ and the materials chemical composition, nor their irradiation conditions.

In Fig 1a, a trend of $N$ to increase with the received neutron dose can be made out, despite the scatter. However, at constant dose, $N$ varies within more than an order of magnitude of values. This is intrinsically understood as the consequence of the varying chemical compositions between the different materials included in each data set. For what concerns the average diameter $D$, however, Fig. 1b shows no discernible trend with the received neutron dose. Overall, $D$ varies in the range $D \approx 1 - 4$ nm. Like $N$, one would expect $D$ to be influenced by the chemical composition of the material, and therefore to vary in a wide range within the variety of materials included in Tab. 1. Intriguingly, however, a closer look reveals that this variation is due to the fact of plotting together different datasets. When taken separately, $D$ within the same set varies within a narrow range of values, which even appear to be independent of composition, in addition to dose. Indeed, $D$ varies in a limited range in Almirall's set [30] and Jenkins 'set [33], despite the widely different chemical compositions of their materials. Likewise, $D = 3$ nm for all cases, but one ($D = 4$nm), in Huang and Auger's set [31,32], which contains two surveillance steels irradiated in a wide range of doses (from 0.03 dpa to 0.28 dpa). Similar observations hold comparing the studies of Miller [41], Courilleau [40] and Kuleshova [42-44]: they all studied nearly identical Russian-type (VVER) materials but they reported different diameters. The $D$ reported is almost constant with dose in all data sets, with only a slight tendency to increase; this remains remarkably true for the Courilleau set, where irradiation occurred at two different temperatures (265°C and 300°C). These observations strongly suggests that the value of $D$ is significantly influenced by the post-experiment data analysis approach specifically performed in each laboratory. Outstandingly, this influence seems in fact to dominate and to shadow the effects that one would otherwise expect, such as the effect of chemical composition and the effect of the received neutron dose.

In this work, we investigate the origins of the above-discussed inter-laboratory differences in the APT data analysis. As tools to conduct this investigation, we combine two complementary theoretical models:

a)  A state-of-the-art model strongly rooted in physics for neutron irradiation in ferritic steels [4,5], to produce realistic estimates for reference atomic configurations, i.e., before the APT measurement is performed.

b)  An adequately simplified model for APT measurement, modifying the reference atomic configuration to mimic the processes of ions evaporation from the sample tip, their flight towards the detector, and the subsequent image reconstruction. We then apply a standard algorithm for the identification of clusters of the alloying elements in the reconstructed image, from which the N and D descriptors are deduced.

By matching our theoretical predictions with those reported experimentally, we show that the difference between all data sets in Fig. 1 is primarily due to the inconsistency of the data processing between laboratories. This is done by looking for the set of parameters in our theoretical models that provide best agreement for $N$ and $D$ with the experimental values. In this way, we can quantitatively understand why each APT result differs from the others. This procedure enables us to study in detail how the choice of the parameters of analysis influences the reported $N$ and $D$, thereby revealing possible laboratory or technique dependent biases.

This paper is organized as follows. In section 2, we describe the method we employed to simulate the generation of APT-like data, and the subsequent cluster identification, resulting in the deduction of $N$ and $D$, for any given sample of RPV material. The focus is on the identification of which parameters are intrinsically involved and are thus expected to vary between different measurements. In section 3, we report how the choice of the parameters identified in section 2 influences the density and size of the observed population of solute clusters. Last, we show in section 4 that, if we analyse the reference microstructures (from the model) that correspond to each experimental data point, using the same analysis method and parameter set, the scatter in Fig. 1b disappears, and trends become much clearer.

## 2 Method: theoretical simulation of an APT experiment

Our method to simulate an APT experiment is depicted in Fig. 2. It proceeds in four mains steps, which are briefly summarized here and described in detail in the following subsections. Within these steps, a total of 18 parameters are defined, as summarized in what follows (all these parameters will be better explained in the subsequent sections). Given a material and its irradiation conditions:

- **In step 1** (described in section 2.1), we provide a reference microstructure in the material after irradiation to the prescribed neutron dose, under given conditions, before the APT measurement is performed. We employ an object kinetic Monte Carlo (OKMC) model, which requires 11 input parameters, namely the concentrations of solute atoms ($C_{Ni}$, $C_{Mn}$, $C_{Si}$, $C_P$, $C_{Cu}$, $C_{Cr}$, $C_C$), the irradiation temperature, $T_{irr}$, the neutron dose rate, $R_{irr}$, the average grain size, $D_{GB}$, and the dislocation density, $\rho_d$.

- **In step 2** (described in section 2.2), we construct a detailed atomic configuration out of the coarse-grained estimate derived in Step 1. This atomic configuration (Fig. 2a and Fig. 2b) is used as reference configuration before the APT measurement is performed. A key parameter, $C_{Fe}$, is defined in this step, which is related to the *a priori* unknown Fe content in the solute clusters.

- **In step 3** (described in section 2.3), we generate an APT-like configuration (Fig. 2c), i.e., an estimate of the 3D atomic configuration that is reconstructed during the measurement out of the evaporated reference configuration obtained in step 2. Three parameters are introduced at this step, namely: the single ion detection efficiency, $P_{APT}$, the corrected lattice parameter, $A_0$, and the anisotropic lattice distortion, $M_{APT}$.

- **In step 4** (described in section 2.4), we apply a standard algorithm (DBSCAN [47]) to identify clusters of solute atoms in the APT-like configuration (Fig. 2d), thus determining both the number density $N$ and the average diameter $D$, which conclude the APT measurement. Three more parameters are defined in this step, namely: the number of neighbours for belonging to a cluster, $KNN$, the maximal distance between pairs in the cluster, $d_{max}$, and the minimum number of atoms in a cluster, $N_{Min}$.

The 11 parameters related to the OKMC simulations ($C_{Ni}$, $C_{Mn}$, $C_{Si}$, $C_P$, $C_{Cu}$, $C_{Cr}$, $C_C$, $T_{irr}$, $R_{irr}$, $D_{GB}$ and $\rho_d$) are imposed by the materials studied in this work, and their irradiation conditions (see supplementary information); they are thus not investigated in this work. However, the other 7 parameters ($C_{Fe}$, $P_{APT}$, $A_0$, $M_{APT}$, $KNN$, $d_{max}$ and $N_{Min}$) are related to the simulation of an APT measurement, up to the calculation of the number density $N$ and average diameter $D$ of the solute clusters. They are thus the focus in this work.

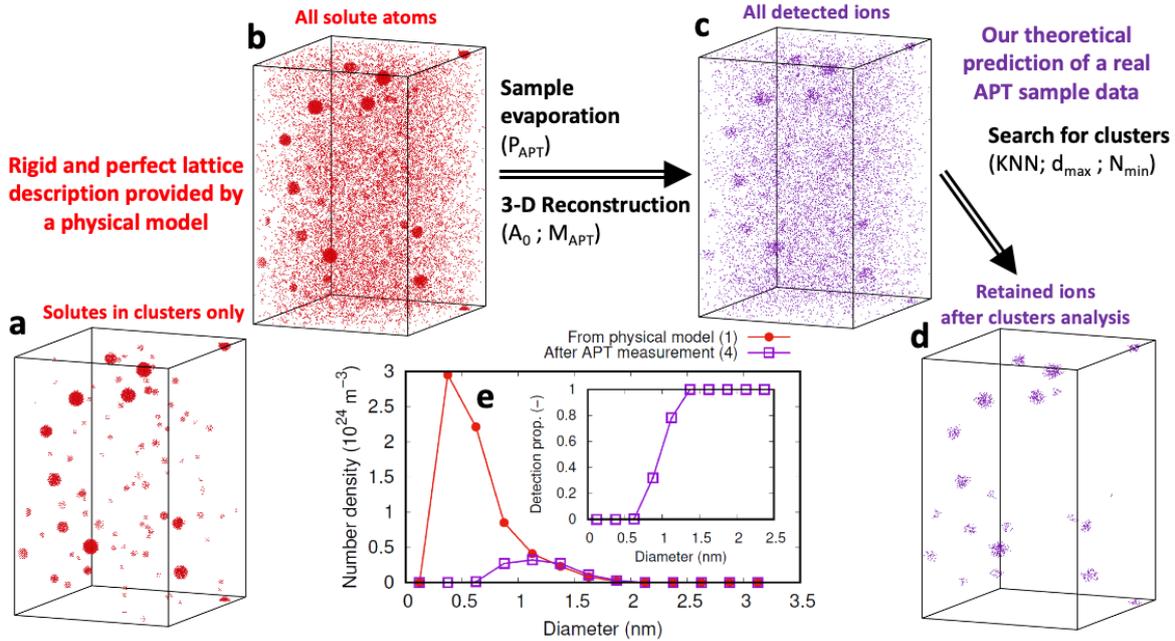

**Fig. 2 –** Method used in this work for the theoretical simulation of the process of sample analysis with the APT technique, here illustrated for the "R39" material in the Almirall set, assuming $C_{Fe}$ = 0.5, $P_{APT}$ = 0.5, $A_0$ = 0.287nm, $M_{APT}$ = 1nm, $KNN$ = 1, $d_{max}$ = 0.43nm, $N_{Min}$ = 20. Atomic structures here contain no more than 1.2 million atoms before evaporation for the convenience of illustration. They are smaller than the simulation boxes used for this study, which contained 9.1 million atoms before evaporation. **(a)** Perfect-lattice atomic configuration provided by the physical model (only the solute clusters are displayed); **(b)** Same as (a), but solute atoms still dissolved in the matrix are also displayed; **(c)** The evaporation of the sample, and the reconstruction of a 3D atomic structure, are emulated; **(d)** Clusters of solute atoms are identified in the reconstructed atomic structure; **(e)** Comparison of the solute clusters size distributions before and after the APT measurement. The inset graph shows, for each cluster diameter in the source configuration, what proportion is detected after cluster analysis.

## 2.1 Estimation of the initial microstructure with a coarse-grain microstructure evolution model.

However powerful, APT does not provide a fully faithful and complete representation of the atomic structure in the material under exam. APT implies evaporation of a needle-shaped sample, followed by a reconstruction step (see later in <span style="color:red">section 2.3</span>): detailed configurations at sub-nm scales are affected in a destructive way and cannot be reverted. Therefore, in the absence of more detailed information coming from direct experimental evidence, and in the absence of more elaborate models, in the past the influence of the intrinsic parameters of APT analysis was checked on artificially constructed reference atomic configurations, assuming simplified populations of solute clusters, see e.g. <span style="color:red">[25, 26]</span>.

In this work, we make use a theoretical model to construct reference atomic configurations that does not make direct assumptions regarding how the solute clusters would be found before APT analysis. For this, we rely on a recently developed microstructure evolution model strongly rooted on physics <span style="color:red">[4,5]</span> for neutron irradiation in RPV steels. The model is based on the object kinetic Monte Carlo (OKMC) method to simulate the detailed evolution of radiation-induced defects and subsequent formation of solute clusters as a function of neutron dose, for a given irradiation temperature and dose rate, in a small simulation volume that is representative of any location in a grain of the alloy. A full description of the OKMC model goes beyond the scope of this paper: interested readers are referred to <span style="color:red">Ref. 4</span> and <span style="color:red">Ref. 5</span> for a detailed description. The model can be concisely summarized as follows:

- The materials are described by:

  o Their chemical composition: $C_{Ni}$, $C_{Mn}$, $C_{Si}$, $C_P$ and $C_{Cu}$ are the solute atom contents, given in atomic percent. The simulation starts with an initial random solution of the solute atoms in the otherwise perfect bcc matrix with all the other sites occupied by Fe atoms.

  o C atoms as impurities are not explicitly added to the chemical composition vector, because the explicit treatment of C migration is known to slow down enormously the simulation. Nevertheless, C impurities are introduced in an effective way in the relevant concentrations $C_C$, via trap events for migrating point defects.

  o Cr solute atoms are generally not expected in significant amounts in RPV and model alloys materials ($C_{Cr}$ < 0.2%), but they are found ($C_{Cr} \approx 2.5\%$) in VVER materials. Like in Ref. [48], we assume that Cr has no interaction with vacancies, nor with the other solute atoms, but does influence the migration of self-interstitial atoms (SIA). In practice, Cr atoms are not explicitly introduced in the simulation volume. Instead, we apply the "grey alloy" approximation model proposed in Ref. [48] and we adapt the attempt frequency for SIA defects migration as a function of the $C_{Cr}$ content in the material.

  o Large and non-radiation-related features in the microstructure, such as grain boundaries and the network of dislocations, are too large (> μm) compared to the typical size of the OKMC simulation volume (< 100 nm) to be included in it. They are thus indirectly introduced, in terms of probabilistic sink events, calculated according to the corresponding sink strengths from the rate theory [49]. These depend explicitly on the average grain diameter $D_{GB}$ (in micrometres), and the density of the network of dislocations $\rho_d$ (in m$^{-2}$), which are thus the input given to the OKMC simulation.

  The above-listed variables are enough to describe and adequately differentiate the different kind of materials of relevance for this study, i.e., RPV steels, their VVER counterparts and model alloys.

- The irradiation conditions are specified by the temperature $T_{irr}$, and the desired dose rate $R_{irr}$ (typically expressed in displacement per atoms per seconds, noted as dpa/s). Irradiation events are simulated by introducing atomic collision cascades at a given rate, picked up from a database depending on the desired dose rate. These cascades were simulated separately by molecular dynamics and only the debris lefts in the material after the cascades cooled down is stored in libraries. In practice, in the OKMC simulation, when an irradiation event is selected, vacancy and SIA defects are added to the simulation volume as specified in the cascades debris libraries, selecting energy values that provide the correct dose increment, in terms of displacements per atoms (dpa), for the target dose-rate.

- Thermally activated events are stochastically chosen following the OKMC algorithm, using as probability their frequency expressed according to the transition rate theory. They account for migration events or dissociation events affecting a vacancy or an SIA, and their clusters, containing solute atoms as well. The most essential events worth emphasizing here are:

  o Solute atoms are allowed to form stable pairs with either a single vacancy or a single SIA defect: once formed, the pair has a migration and a dissociation event as a single defect. Consequently, the solute atom is dragged by the point defect, until a dissociation event is selected, or the point defect is absorbed by another defect (direct coalescence).

  o SIA defects are assumed to be strongly bound to solute atoms decorating them. In practice, at the temperatures relevant for RPV materials (around 300°C), SIA defects are thus fully immobilized as soon as they are decorated by a couple of solute atoms.

o Solute clusters are allowed to be dissolved by single vacancies only. A key development in the work detailed in Ref. [5] was to propose a general formula to estimate the binding energy between a single vacancy, or a vacancy-solute pair, and the solute clusters. This general formula, inspired from evidence from first principles calculations (DFT), was formulated as a function of the proportion of Ni in the chemical composition vector.

The microstructure evolution predicted by the OKMC simulation is fully detailed and no radiation defects are excluded. Indeed, the internals of the model comprise a list of defects that includes single vacancies, single SIAs, clusters of vacancies, clusters of SIAs, and, most importantly, clusters of solute atoms which are the focus of this work. In Ref. [4,5] we demonstrated that the solute clusters dominantly nucleate on small, immobilized SIA clusters (small dislocation loops of interstitial nature). The model originally proposed in Ref. [4] was limited in predictive capacities, because the key parameters describing the thermal stability of solute clusters ought to be empirically fitted case by case (i.e. for each material being studied) to get a reasonable agreement between the OKMC predictions and the APT experimental measurements. This aspect of the model was significantly improved in Ref. [5], where we proposed a general formula to estimate the missing parameters directly from the chemical composition of the materials, specifically, the relative Ni content. Thus, here, we employ the improved model proposed in Ref. [5], which is fully parameterized and does not require any empirical fitting versus experimental data, using the input parameters listed above. In Ref. [5], the model was applied to a database of 426 materials varying in a wide range of chemical compositions, from typical surveillance RPV materials to chemically tailored model alloys for steels. We showed that, thanks to a reasonably correct prediction of the square root of the volume fraction $\sqrt{f}$ of the solute clusters from the model, it was possible to provide high fidelity assessments of the measured radiation-induced embrittlement, in a very large number of cases (almost 2000 experimental data-points of reference). This demonstrated that the parameterization of the microstructural model is suitable to describe any RPV material, free of assumptions regarding the microstructure after irradiation, i.e., the model is exempt from any arbitrary assumption or fitting towards experimental data. In other words, the model provides a microstructure of reference that is an as-good-as-currently-possible representation of the real microstructure in irradiated RPV steels.

However, the output of the OKMC simulation is a coarse-grained description of the microstructure. Specifically for solute clusters, the information provided is the position (x, y, and z) and the number of solutes in each cluster. The latter, denoted as $n_{Sols}$, is the cumulated number of Ni, Mn, Si, P and Cu atoms found in the cluster. This information cannot be directly used to simulate an APT measurement. In the next section, we describe how the coarse-grained microstructure can be converted into an atomic configuration, to which the same analysis as in an APT experiment can be applied.

## 2.2 Construction of an atomic reference configuration

Fig. 3 depicts the process we employed to convert the coarse-grained description of the material microstructure into an atomic configuration representative of the microstructure before an APT measurement.

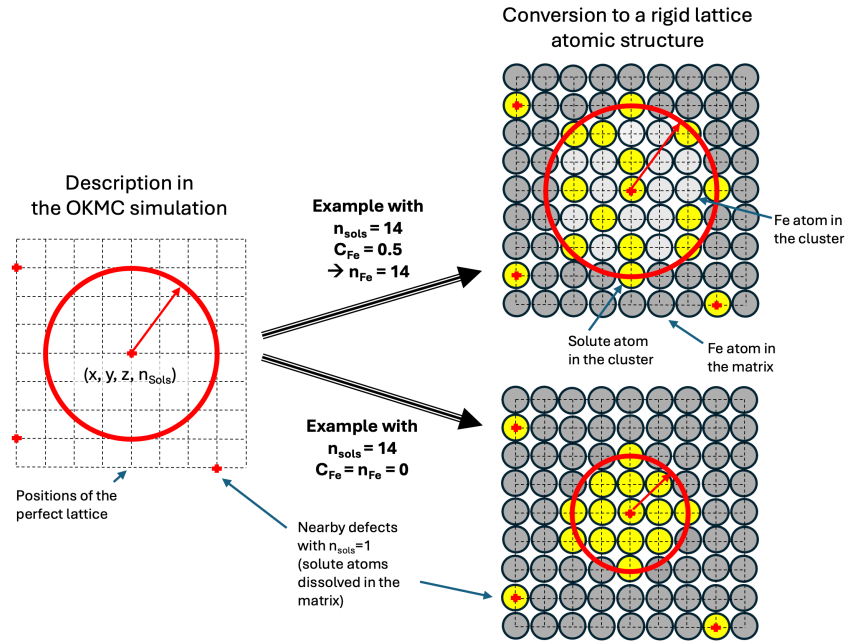

**Fig. 3** – Pictorial representation (in two dimensions) of the construction of a rigid lattice atomic structure for a cluster of solute atoms. The dashed lines represent the grid of the perfect lattice (bcc structure in this work). **Left:** the cluster is described in the OKMC simulation as its centre-of-mass position (x, y, z coordinates), the number $n_{Fe}$ of Fe atoms, and the number $n_{Sols}$ of solute atoms (Ni, Mn, Si, P, and Cu) contained in the cluster. The radius of the defect corresponds to the one that encompasses at least $n_{Fe}+n_{Sols}$ atomic sites. **Right:** All lattice sites are filled with an atom. The prescribed numbers of Fe and solute atoms are placed at randomly chosen sites within those close enough to the centre of mass. Two examples are depicted for illustration of limiting cases: one case where $C_{Fe} = n_{Fe}/(n_{Fe}+n_{Sols}) = 0.5$, and another case where $C_{Fe} = n_{Fe} = 0$.

On the left side in Fig. 3, an individual solute cluster is located at a given lattice site (x, y, z coordinates) in the simulation box. The cluster is described by $n_{Sols}$ and by the number of atoms of each chemical species. Unfortunately, however, because of its coarse-grained nature, the model provides no information about how the $n_{Sols}$ solutes atoms are truly distributed in space around their centre of mass. During an OKMC simulation, it is implicitly assumed that all clusters are compact spheres, which is an instrumental hypothesis to define the capture radius of the defects (red arrows in Fig. 3). This is of course a strong assumption, for at least two reasons:

A. Non-spherical, but rather disk-shaped solute clusters are sometimes observed during APT investigations of irradiated steels. This is especially true for ferritic-martensitic steels irradiated to high dpa [50], but they have also been reported for RPV and VVER steels [4,5,51,52]. Such disk-shaped clusters are likely to be direct evidence of solute decoration of small SIA loops, which cannot be directly confirmed with APT investigations alone.

B. Aside from their shape, the determination of the true Fe content in solute clusters as detected by APT is a long-standing debate in the nuclear materials community. When this information is reported, clusters are typically dilute and contain 50%-80% Fe (see references cited in Tab. 1). According to some, such a high iron content should be an artefact of the atom probe technique [11,53,54]. Recently, Courilleau et al [40] analysed this aspect in detail and reported convincing evidence that the Fe content was as high as 75%-80% in their samples and that this was most likely not an artefact. Other techniques (electron microscopy, small angle neutron scattering [11,55-57] suggest contents round 20% Fe. From the standpoint of the OKMC model, an estimation of the predicted Fe content was given is Ref. [5] to be around 50%. This number was based on the accumulated counts of single SIA defects, free of solutes, absorbed by the solute cluster, which were assumed to bring one Fe atom to the cluster. The presence of Fe atoms leads to a dilute and diffuse distribution of solutes, which may depart significantly from a perfect sphere.

However, the occurrence of ring-shaped solute clusters, although documented, remains a rare finding. On the other hand, the analysis of APT reconstructions eventually always associates a diameter with visible clusters, thus implicitly assuming that these can be brought down to a roughly spherical shape. Therefore, we assumed in this work that all solute clusters have a perfect spherical shape in the reference microstructure. The number $n_{Sols}$ of solute atoms to be distributed inside the sphere is taken from the OKMC model. In addition, even though the model does provide information about the Fe content of clusters [5], we deliberately decided to consider the actual Fe content of the clusters as an unknown, and to vary it, to investigate limiting cases. For this reason, we impose the parameter $C_{Fe}$ as the Fe content in the solute clusters and study its influence on the best-match analysis parameters. Given $C_{Fe}$ and $n_{Sols}$, we calculate the number $n_{Fe}$ of Fe atoms that belong to a solute cluster as:

$$n_{Fe} = \frac{C_{Fe}}{1 - C_{Fe}} n_{Sols} \tag{1}$$

On the right-hand side in Fig. 3, the specified numbers of solute and Fe atoms are placed at random positions (random solid solution) within a sphere centred on the OKMC-specified centre-of-mass position, with a radius calculated as:

$$R = a_0 \left( \frac{3(n_{Sols} + n_{Fe})}{8\pi} \right)^{1/3} \tag{2}$$

Here, $a_0$ is the lattice parameter, assumed to be 0.287 nm in this work. To explore limiting cases, we considered the following values for the $C_{Fe}$ parameter:

- **$C_{Fe}$ = 0** corresponds to the hypothetical situation where solute clusters are free of any significant content of Fe, which is consistent with the late blooming phases [2] hypothesis that they are seeds for the formation of thermodynamically-stable foreign phases.

- **$C_{Fe}$ = 0.5** is to be considered as the lower range of APT-reported values. It also corresponds to the value theoretically predicted by the model [5], as explained above.

- **$C_{Fe}$ = 0.75** is to be considered as the upper range of APT-reported values [40].

## 2.3 Generation of APT-like atomic configurations.

This step corresponds to the transition between Fig. 2b and Fig. 2c. The APT measurement consists of a controlled atom-by-atom erosion from a tip-shaped sample, the detection of the evaporated ions after their flights towards a detector, and, finally, the reconstruction of a three-dimensional image. These processes are commonly referred to as *field evaporation*. A review of modelling methods for field evaporation can be found in Ref. [58]. These complex approaches imply modelling the dynamic evolution of the sample tip under the application of high electric tension and/or laser pulses—accounting for several factors such as the local lattice structure and orientation, surface migration of atoms, etc.—and modelling the trajectories of evaporated ions towards the detector. In this work, we adopted a simpler method to model the generation of APT-like atomic configurations, inspired from our past work in Ref. [25], where these processes are globally accounted for with no more than 3 variable parameters, as depicted in detail in Fig. 4:

- **$P_{APT}$** stands for the single-ion detection efficiency of the atom probe. As we see in Fig. 4b, each atom of the simulated system is either considered as 'detected' or as 'undetected' and therefore removed. In this work, we assume that $P_{APT}$ = 0.4, 0.5 or 0.6, which realistically represents most APT setups used around the world.

- **$A_0$** is the corrected lattice parameter, to account for isotropic distortions of the reconstructed volume due to an imprecise estimation of the volume of the evaporated sample. Indeed, the reconstruction process is not uniform across different laboratories. The best practice is to calibrate the reconstructed volume comparing scanning or transmission electron microscopy images of the tip made before and after the APT experiment. Due to limited time and resources, many operators use other techniques including calibration through crystallographic poles [59] or assuming constant density – all of which contributes to significant imprecision in the reconstructed sample volume [60]. Small changes in reconstruction parameters, such as the tip profile (radius and shank), electrostatic field factor and evaporation field, fixed image compression factor, etc., may result in large length-scale variations. $A_0$ may thus be smaller, equal, or greater than the lattice parameter $a_0$ = 0.287nm we usually consider as being the average one for RPV materials (see Eq. (2) here and Ref. [4,5]). At this stage (Fig. 4c), the atomic structure is still a rigid lattice, being homogeneously swollen or shrunk compared to the atomic structure in Fig. 4b. We assumed $A_0$ = 0.2575nm – 0.3616nm, which correspond to an isotropic volume change of the crystal by a factor $(A_0/a_0)^3 = 0.72 - 2$.

- **$M_{APT}$** accounts for the anisotropic lattice distortions induced in the reconstruction by the evaporation and detection process. For simplicity, our model does not include trajectory aberration effects, which are likely induced by the differences in ion evaporation energies between atoms located in a cluster of solutes, and those located in the matrix near the cluster. The consequence of such aberrations can be to induce an apparent increased concentration of Fe in the clusters of solute atoms being identified in the reconstructed atomic configuration. Likewise, our model does not include surface migration effects, which would form crystallographic poles, hence varying the density and noise in some XY coordinates of the reconstructed configuration. Like in our work of Ref. [25], each atomic coordinate is shifted by a random distortion vector, the module of which is chosen following a Gaussian distribution, with full width at half maximum (FWHM) denoted as $M_{APT}$. To account for the anisotropy of the evaporation of the needle-shaped sample, we define $M_{APT}^{(X,Y)}$ as the distortion applied in the XY plane (assumed to be perpendicular to the needle axis), while $M_{APT}^{(Z)}$ is applied on the Z direction (assumed to be aligned with the needle axis). In Ref [25], we established that $M_{APT}^{(Z)}$ should be as small as 0.1 nm, while $M_{APT}^{(X,Y)}$ may be larger in magnitude. In this work, we made vary $M_{APT}^{(X,Y)}$ = 0.5 nm, 1 nm, or 1.5 nm, consistently with the findings in Ref. [25, 27].

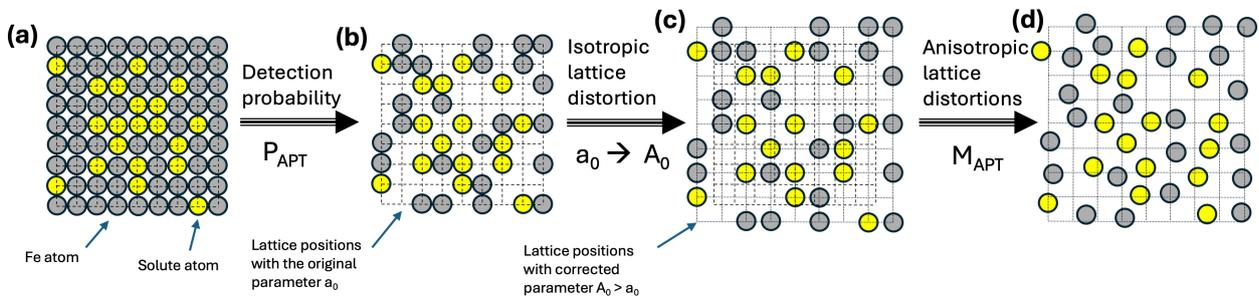

**Fig. 4** – Pictorial representation (in 2 dimensions) of the three steps followed to generate an APT-like atomic configuration, given as input a reference atomic configuration.

## 2.4 Search for clusters of solute atoms

This step corresponds to the transition between Fig. 2c and Fig. 2d. To identify clusters of solute atoms in the atomic structure depicted in Fig. 2c and in Fig. 4d, which corresponds to the reconstruction of an atom probe experiment, we employ the density-based spatial clustering of applications with noise (DBSCAN) method, as described in Ref. [47]. The variable parameters associated with this method are:

- $d_{max}$ is defined as the maximal distance allowed between pairs of atoms so that they are considered as close neighbours. In this work, we assumed $d_{max}$ = 0.43nm, 0.57nm, 0.65nm or 0.72nm.

- **The $K$ order** (also noted as **KNN**) is an integer number which defines how many close neighbours are required for an atom to be considered as a *core atom*. Atoms which count less than $K$ close neighbours, but at least one core atom, are considered as border *atoms*; the other atoms are considered as *noise atoms*. In this work, we do not vary this parameter, and we assume KNN = 1. Our cluster analysis method is thereby equivalent to the maximum separation method (MSM), which was used by most APT authors cited in Tab. 2.

- $N_{Min}$ is defined as the minimal number of atoms (core + border) that clusters should count. Clusters with less than $N_{Min}$ atoms are disregarded, and their atoms are labelled as noise.

After the clusters' identification, the number density $N$ is calculated as the total number of clusters divided by the volume of the simulation box (calculated using $A_0$, and not $a_0$). The average cluster diameter $D$ is the average of all individual cluster diameters, taken to be twice the gyration radius.

It is worth noting that two additional steps customarily exist in the analysis of experimental APT data. An "enveloping" step first adds to every cluster any noise atom sufficiently close to any of its components. The purpose of this enveloping step is to ensure that individual atoms inside the clusters were not wrongly labelled as noise (one might expect such limiting cases if the local density is low). The drawback of the enveloping step is that a shell of noise atoms is systematically added to the outer surface of the cluster. An "erosion" step is therefore applied as compensation. However, since in this work the clusters are from the start constructed with spherical shape (see Fig. 3), the enveloping-erosion steps are unnecessary. We thoroughly verified this in suitable limiting cases. Therefore, for the sake of simplicity, and for the sake of not introducing extra variable parameters (enveloping and erosion distances) to this study, the enveloping-erosion steps were not applied.

Finally, it is also worth noting that all the steps described in section 2 work with a computational sample volume where periodic boundary conditions are applied in the X, Y and Z directions. This feature is inherited from the OKMC simulation. In contrast to real APT experiments, therefore, we do not have issues with features being partially cut by the limits in the field-of-view of the microscope, which contribute to extra uncertainties to the reported $N$ and $D$ [61].

# 3 Results

OKMC simulations were performed for the 74 cases (varying materials and irradiation conditions) listed in Tab. 1; the detailed information about the input variables $C_{Ni}$, $C_{Mn}$, $C_{Si}$, $C_P$, $C_{Cu}$, $C_{Cr}$, $C_C$, $T_{irr}$, $R_{irr}$, $D_{GB}$ and $\rho_d$ for each individual simulation is given in Tab. SI1-SI3 in the supplementary information. We used a simulation box with dimensions 150 $a_0$ × 160 $a_0$ × 190 $a_0$, which contained 9.12 million atoms before evaporation, and thus roughly $3.7 - 5.5$ million atoms after evaporation, depending on the $P_{APT}$ parameter. The minimal defects density (1 defect in the simulation box) is thus $4.6 \cdot 10^{21} - 1.3 \cdot 10^{22}$ m$^{-3}$ depending on the $A_0$ parameter. The Z axis is assumed to be aligned with the axis of the APT needle sample. In each case, the APT emulation procedure as described in section 2 was first applied varying the values of the parameters, to see their effect on the outcome of the analysis, with respect to the reference microstructure, known from the OKMC model.

Our analysis reveals (Fig. 5) that, at constant $A_0$, the parameters $d_{max}$ and $N_{min}$ have the strongest influence on the resolved cluster size distribution, while $P_{APT}$ and $M_{APT}$ have secondary impact in most cases, even if varied through a wide range. Interestingly, we see that clusters that are known to exist in the source configuration may be either detected, or not detected, after APT analysis, with a probability that is mainly determined by $N_{Min}$. Too small values such as $N_{Min} \leq 5$ may result in the wrong identification of clusters composed of atoms that are truly dissolved in the matrix [62]. We see that, with $N_{Min} \geq 10$, it becomes difficult to tell below which threshold size the APT cannot detect clusters; instead, there is a range of sizes ($D \leq 1.5$ nm) where clusters may or may not be detected, the probability of detection being mainly determined by the combination of $P_{APT}$ and $N_{Min}$. In this range, this probability gradually increases from 0 to 1 as a function of increasing cluster size. When detected, the resolved solute clusters may appear similar, or not, to the original cluster in the source configuration. If $d_{max}$ is small enough, the detected clusters tend to appear smaller, or at most of identical size (Fig. 5a), with respect to the source configuration. If, on the contrary, $d_{max}$ is too large, we see (Fig. 5b) that the whole size distribution is shifted to the right, as a direct consequence of the systematic integration of nearby solute atoms that were originally still dissolved in the matrix. Other undesired effects also happen: (a) clusters that are originally separate entities appear as one (Fig. 5ef); (b) clusters may be formed out of atoms that were originally dissolved in the matrix (not depicted in Fig.5).

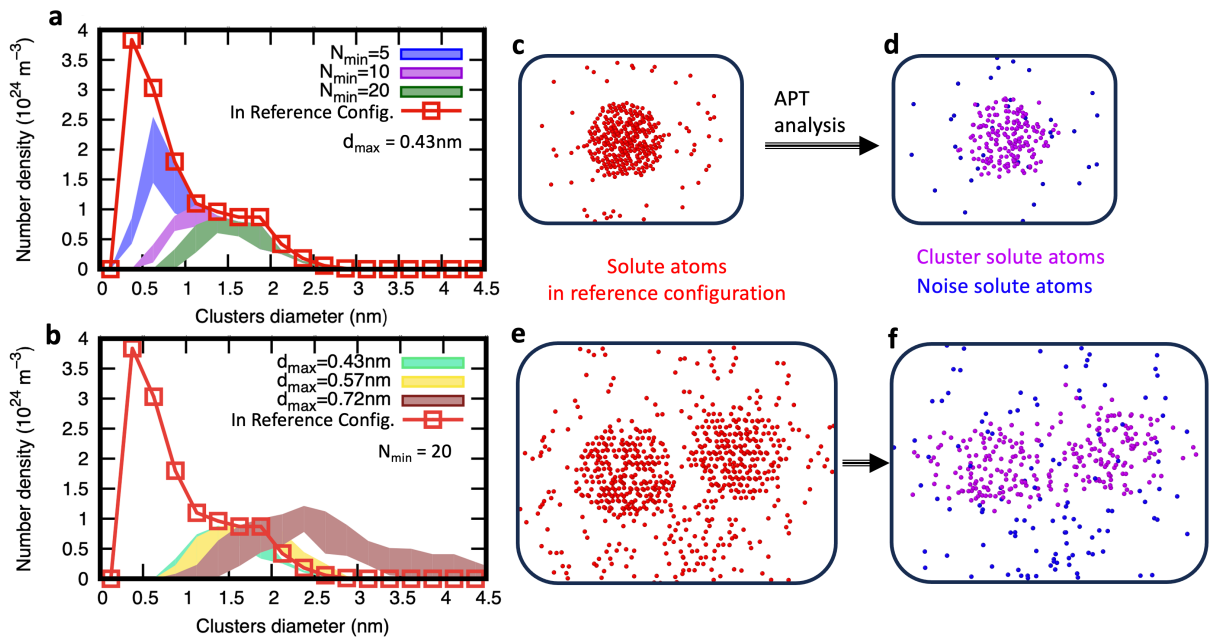

**Fig. 5** – Application of the APT emulation algorithm (section 2.3) to a reference atomic structure constructed from the OKMC predicted microstructure (section 2.2), for the "R22" material of the Almirall set [30], chosen here as an example. **(a, b)** Size distributions for the detected solute clusters, compared with the size distribution in the source configuration. The coloured bands show the range of variation when $0.4 \leq P_{APT} \leq 0.6$ and $0.5$nm $\leq M_{APT} \leq 1.5$nm are varied, with $A_0$, $d_{max}$ and $N_{Min}$ kept constant to the stated values. In all cases, KNN = 1 and $C_{Fe}$ = 0.50; **(c,d)** Ideal situation when a cluster in the source configuration is detected as a similar feature after APT analysis. Only the defect apparent diameter may be affected by $d_{max}$, which influences how many

solute atoms in the nearby matrix are assimilated to the cluster; **(e,f)** Example of non-ideal situation when two clusters which are separate features in the source configuration are seen as one feature after APT analysis.

One of the main objectives of this work was to use the methodology of section 2 to investigate quantitatively the origin of inter-laboratory differences (Fig. 1b) in the APT data analysis. For this purpose, we rely to our OKMC model to provide estimates of the initial microstructures before the APT measurement is performed. Then, the best-match parameters between our predicted values of $N$ and $D$ and those reported experimentally is looked for, by varying the analysis parameters. Given the results shown in Fig. 5, we considered $P_{APT}$ and $M_{APT}$ as sources of noise, and thus these parameters were only varied to estimate the uncertainty they induce (bands instead of lines). Namely, all simulations of APT measurement were performed 9 times, with the combinations of $P_{APT}$ = 0.4, 0.5 or 0.6 and $M_{APT}$ = 0.5 nm, 1 nm or 1.5 nm. Since we assume KNN = 1, only the $C_{Fe}$, $A_0$, $d_{max}$ and $N_{Min}$ parameters needed to be varied to find the best match. Considering the discussions in section 1 on the findings presented in Fig. 1b, our goal was to determine the best-match parameters for each of the 11 groups of authors defined in Tab. 1 individually, and independently of each other. We must however discuss what parameters should be varied for each individual groups, or not:

- $C_{Fe}$, the concentration of Fe in the solute clusters in the reference atomic configurations, is imposed. As discussed in section 2.2, we considered $C_{Fe}$ = 0, 0.5 or 0.75. This parameter represents an uncertainty during the microstructure evolution under neutron irradiation, i.e., before the APT measurement is performed. We guess that $C_{Fe}$ may not be a constant value, influenced as it is by the materials composition and the irradiation conditions. However, for the sake of simplicity, in this work we assumed $C_{Fe}$ to be identical for all materials in all irradiation conditions. Therefore, $C_{Fe}$, when varied, was interpreted as a global parameter, applied identically to all 74 cases, regardless of which the 11 groups they belong.

- On the contrary, the $A_0$, $d_{max}$ and $N_{Min}$ parameters are those that need to be tuned to describe each of the 11 groups individually. Of the three, $A_0$ is connected to the accuracy of the estimation for the evaporated volume of the sample (section 2.3), while $d_{max}$ and $N_{Min}$ are deliberate choices made by each researcher during the identification of clusters (section 2.4). In this work, we assumed that each of the 11 research teams defined in Tab. 1 treated all their own measurement data in an identical manner, and therefore that a unique value of $A_0$, $d_{max}$ and $N_{Min}$ can be used to simulate all their APT measurements, for all cases within their group. We considered all the combinations of the following values:

  o $A_0$ = 0.2575nm, 0.2722nm, 0.2870nm, 0.3056nm, 0.3243nm or 0.3616nm, which correspond to an isotropic volume change of the reference atomic configuration by a factor $(A_0/a_0)^3$ = 0.72, 0.85, 1, 1.21, 1.44 or 2, respectively.

  o $d_{max}$ = 0.43nm, 0.57nm, 0.65nm or 0.72nm

  o $N_{Min}$ = 5 − 140

Given the strategy described above, we performed a total of 47 952 simulations of APT measurements; this number is the result of 74 cases × 3 values for $P_{APT}$ × 3 values for $M_{APT}$ × 3 values for $C_{Fe}$ × 6 values for $A_0$ × 4 values for $d_{max}$, while any value for $N_{Min}$ can be applied afterwards. This number motivated the size of the OKMC simulation box (150 $a_0$ × 160 $a_0$ × 190 $a_0$), which was estimated as the largest to allow the 50 544 simulations of APT measurements to be conducted in an affordable computing time. We then performed 3 × 11 = 33 parameters optimizations, as reported in Tab. 2 and as illustrated in Fig. 6. During each optimization, $C_{Fe}$ was fixed, and we found which combination of the $A_0$, $d_{max}$ and $N_{Min}$ parameters maximized the agreement between the values of $N$ and $D$ reported by the experimental works, and our predicted values. We minimized the residual error, calculated as:

$$E(A_0, d_{max}, N_{min}) = E_N(A_0, d_{max}, N_{min}) + E_D(A_0, d_{max}, N_{min}) \tag{3}$$

Where,

$$E_N(A_0, d_{max}, N_{min}) = \sum_{P_{APT}=0.4,0.5,0.6}^{\square} \sum_{M_{APT}=0.5,1,1.5}^{\square} \sum_{p=1}^{P} \frac{\left| log\left(N_{APT}^{(p)}\right) - log\left(N_*^{(p)}(P_{APT}, M_{APT}, A_0, d_{max}, N_{min})\right)\right|}{25-21} \tag{4}$$

and

$$E_D(A_0, d_{max}, N_{min}) = \sum_{P_{APT}=0.4,0.5,0.6}^{\square} \sum_{M_{APT}=0.5,1,1.5}^{\square} \sum_{p=1}^{P} \frac{\left| D_{APT}^{(p)} - D_*^{(p)}(P_{APT}, M_{APT}, A_0, d_{max}, N_{min})\right|}{5-0.5} \tag{5}$$

Here, $P$ stands for the number of data points in the set, $N_{APT}^{(p)}$ is the number density $N$ reported by the cited APT study for the data point $p$, and $D_{APT}^{(p)}$ is the corresponding average diameter $D$, while $N_*^{(p)}$ and $D_*^{(p)}$ are the values predicted by our model (which are functions of the $P_{APT}$, $M_{APT}$, $A_0$, $d_{max}$ and $N_{Min}$ parameters). We should note that the numerical factors in the denominators in Eq.(4) and Eq.(5), which correspond to the ranges plotted in Fig. 6, are introduced to make both terms in Eq.(3) dimensionless, thus ensuring an equal weight for both fitting criteria.

The best match parameters we found are summarized in Tab. 2. The comparison between the experimental values of $N$ and $D$, and the predicted values by our model after optimization, are shown in Fig. 6. We see in Tab. 2 that, for the Almirall, Jenkins, Courrilleau, and Takeushi data sets, the best match parameters follow as expected a monotonic evolution with $C_{Fe}$: the largest $C_{Fe}$, the more dilute and the bigger (see Eq. (1-2)) are the solute clusters in the reference configuration, before the APT measurement is simulated. Consequently, if $C_{Fe}$ increases, the best-match parameters will naturally lead to a decrease of $d_{max}$, and/or a decrease of $N_{Min}$, and/or decrease of $A_0$ to match the same targeted (fixed) experimental values of $D_{APT}$ in Eq. (5). The other sets show a non-monotonic evolution, before increasing $C_{Fe}$ leads to changes in the best match parameters that do not strictly respect the previously mentioned logic. The only exception is the LongLife set, for which the best match parameters are identical for all values of $C_{Fe}$.

In Fig. 7, the optimal parameters are compared to those employed by the authors of the APT studied, if reported in the references cited in Tab. 1. We see that the parameters are best correlated in the case $C_{Fe}$ = 0.75. In the other cases, namely $C_{Fe}$ = 0.5 or $C_{Fe}$ = 0, our $N_{Min}$ parameters are systematically, and substantially, larger than those reported experimentally. The $d_{max}$ best match parameter is also, in some cases, significantly larger than in the experimental analysis.

Fig. 8 shows how the experimental $D$ values scale with $N_{Min}$ and $d_{max}$, specifically for the case $C_{Fe}$ = 0.75. We see that, unsurprisingly, the values of $D$ reported experimentally tend to increase with increasing $N_{Min}$ and $d_{max}$ parameters used by the authors of the studies. The same correlation is visible between our optimized parameters and the targeted experimental $D$. It is worth emphasizing that the values of $N_{Min}$ used to analyse the experiments do not exceed $N_{Min}$ = 30 in Tab. 2 (at least the value that are reported), while our best match values rise up to $N_{Min}$ = 100. It is of course unfortunate that the authors cited in Tab. 2 did not systematically report their values of $N_{Min}$, as this would have made a detailed comparison possible. Noticeably, the value of $N_{Min}$ was not specified in the Dohi, Kuleshova and Huang-Auger sets, for which the reported $D$ values are the largest (D > 2.5 nm). It can be surmised that, consistently, also the $N_{min}$ value they used was larger than in other cases, but we cannot verify this hypothesis.

Fig. 8 shows the relation between the targeted $D$ and the optimized $A_0$ parameter for our model. We should emphasize that, since this parameter was introduced to account for a calibration uncertainty, there is no experimental analysis parameter with which it can be compared. Nevertheless, there is no clear correlation between $A_0$ and $D$. This suggests that the parameter $A_0$ is indeed accomplishing its intended role in our model, which is to compensate for imprecise calibration of the experimental setups.

Finally, Fig. 9 shows the number of occurrences for each optimized value of $A_0$. For clarity, these are reformulated in the figure as the effective uncertainty on the volume of the evaporated sample. We see that these uncertainties clearly tend to be on the positive side. This means that, to match the experimental measurements, the volume of our reference atomic configurations generally needs to be inflated. This suggests that the volume of the sample as deduced from the reconstruction after evaporation tends to be an underestimate. We should note that the best match values of $A_0$ go as high as $A_0 = 0.31616$ nm if $C_{Fe} = 0$ or $C_{Fe} = 0.5$, which means that the inaccuracy of estimation on the evaporated volume would be as large as 100%. Of course, this prediction is likely far-fetched. It is however interesting to note that the inaccuracy on the evaporated volume only extends up to 44% if $C_{Fe} = 0.75$, which seems to be a more realistic magnitude. In other words, the best match between microstructure evolution model and experiment results from assuming large Fe contents are found in the clusters.

**Table 2 –** (left columns) Summary of the APT setups and data processing parameters reported by the authors of the cited works. Empty cells are due to missing information; (last 3 columns on the right) best match parameters for the clusters analysis determined in this work (minimization of the residual error in Eq. (3)).

| Name | Ref | Reported parameters in the publications of the experimental APT work | | | Best match parameters determined in this work | | |
|---|---|---|---|---|---|---|---|
| | | APT setup | Reconstruction software and algorithm | Parameters of clusters analysis (if specified) | Optimization 1<br>KNN=1<br>$C_{Fe}=0.75$<br><br>Residual errors:<br>N 57%<br>D 0.15nm | Optimization 2<br>KNN=1<br>$C_{Fe}=0.5$<br><br>Residual errors:<br>N 57%<br>D 0.19nm | Optimization 3<br>KNN=1<br>$C_{Fe}=0$<br><br>Residual errors:<br>N 69%<br>D 0.29nm |
| Almirall | [30] | LEAP 3000X HR<br>LEAP 4000X HR | IVAS<br>MSM | $d_{max}$ 0.5 − 0.6 nm<br>$N_{Min}$ 15-30 | $d_{max}$ 0.57 nm<br>$N_{Min}$ 35<br>$A_0$ 0.3243 nm | $d_{max}$ 0.65 nm<br>$N_{Min}$ 70<br>$A_0$ 0.3243 nm | $d_{max}$ 0.72 nm<br>$N_{Min}$ 140<br>$A_0$ 0.3243 nm |
| LONGLIFE | [5] | LEAP<br>EcoWATAP MSM+<br>LAWATAP | GPM 3D<br>MSM isoposition | | $d_{max}$ 0.72 nm<br>$N_{Min}$ 50<br>$A_0$ 0.287 nm | $d_{max}$ 0.72 nm<br>$N_{Min}$ 50<br>$A_0$ 0.287 nm | $d_{max}$ 0.72 nm<br>$N_{Min}$ 50<br>$A_0$ 0.287 |
| Huang, Auger | [31,32] | ECAP + TAP<br>ECOTAP + LAWATAP | | | $d_{max}$ 0.65 nm<br>$N_{Min}$ 100<br>$A_0$ 0.287 | $d_{max}$ 0.72 nm<br>$N_{Min}$ 100<br>$A_0$ 0.3056 | $d_{max}$ 0.72 nm<br>$N_{Min}$ 140<br>$A_0$ 0.287 nm |
| Jenkins | [33] | LEAP 4000X HR | IVAS 3.8.0<br>MSM | $d_{max}$ 0.45 − 0.65 nm | $d_{max}$ 0.57 nm<br>$N_{Min}$ 15<br>$A_0$ 0.3243 nm | $d_{max}$ 0.65 nm<br>$N_{Min}$ 15<br>$A_0$ 0.3616 nm | $d_{max}$ 0.72 nm<br>$N_{Min}$ 25<br>$A_0$ 0.3616 nm |
| Meslin, Lambrechts | [34-37] | ECOTAP | IVAS<br>GPM3D<br>MSM | $N_{Min}$ 10 | $d_{max}$ 0.43 nm<br>$N_{Min}$ 10<br>$A_0$ 0.3056 nm | $d_{max}$ 0.43 nm<br>$N_{Min}$ 10<br>$A_0$ 0.3243 nm | $d_{max}$ 0.43 nm<br>$N_{Min}$ 10<br>$A_0$ 0.3616 nm |
| Edmondson, Wells | [11, 38,39] | LEAP 3000X HR<br>LEAP 4000X HR | IVAS 3.6.8<br>Matlab script<br>MSM | $d_{max}$ 0.5 nm | $d_{max}$ 0.43 nm<br>$N_{Min}$ 25<br>$A_0$ 0.3243 nm | $d_{max}$ 0.43 nm<br>$N_{Min}$ 25<br>$A_0$ 0.3243 nm | $d_{max}$ 0.43 nm<br>$N_{Min}$ 120<br>$A_0$ 0.2722 nm |
| Courilleau (VVER) (**) | [40] | LEAP 4000X HR | Isoposition | $d_{max}$ 0.4 nm<br>$N_{Min}$ 6 − 12 | $d_{max}$ 0.72 nm<br>$N_{Min}$ 15<br>$A_0$ 0.3243 nm | $d_{max}$ 0.72 nm<br>$N_{Min}$ 20<br>$A_0$ 0.3243 nm | $d_{max}$ 0.72 nm<br>$N_{Min}$ 25<br>$A_0$ 0.3243 nm |
| Miller (VVER) (**) | [41] | LEAP | MSM + EM | $d_{max}$ 0.6 nm<br>$N_{Min}$ 10 | $d_{max}$ 0.57 nm<br>$N_{Min}$ 10<br>$A_0$ 0.3243 nm | $d_{max}$ 0.57 nm<br>$N_{Min}$ 20<br>$A_0$ 0.287 nm | $d_{max}$ 0.65 nm<br>$N_{Min}$ 15<br>$A_0$ 0.3243 nm |

| Kuleshova (VVER) [**] | [42-44] | LEAP 4000X HR | IVAS 3.6.12 MSM | | | $d_{max}$ 0.65 nm $N_{Min}$ 40 $A_0$ 0.3056 nm | $d_{max}$ 0.65 nm $N_{Min}$ 70 $A_0$ 0.2722 nm | $d_{max}$ 0.72 nm $N_{Min}$ 50 $A_0$ 0.3056 nm |
|---|---|---|---|---|---|---|---|---|
| Dohi | [45] | LEAP | Recursive search algorithm | $d_{max}$ 0.5 nm | | $d_{max}$ 0.65 nm $N_{Min}$ 100 $A_0$ 0.2575 | $d_{max}$ 0.72 nm $N_{Min}$ 100 $A_0$ 0.287 | $d_{max}$ 0.72 nm $N_{Min}$ 100 $A_0$ 0.2722 nm |
| Takeuchi | [46] | LEAP | MSM | $d_{max}$ 0.7 nm $N_{Min}$ 20 | | $d_{max}$ 0.65 nm $N_{Min}$ 30 $A_0$ 0.3243 nm | $d_{max}$ 0.72 nm $N_{Min}$ 50 $A_0$ 0.3243 nm | $d_{max}$ 0.72 nm $N_{Min}$ 50 $A_0$ 0.3243 nm |

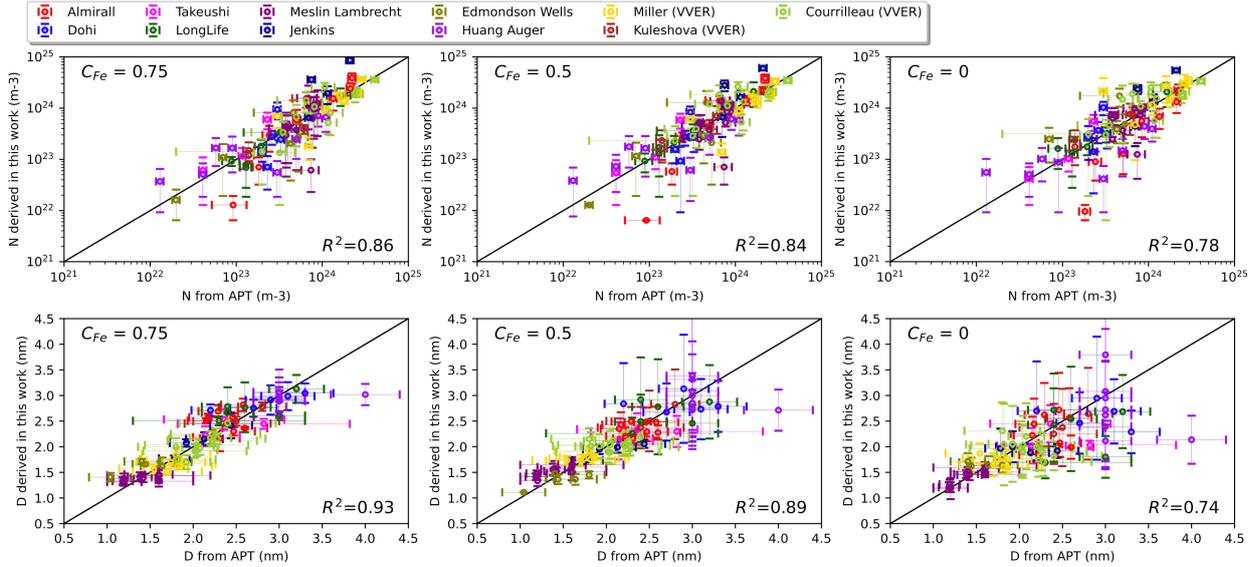

**Fig. 6** – Comparison between *N* and *D* reported in the scientific literature and those derived with the model and method used in this work, after identifying the best match analysis parameters. I.e., for each set defined in Tab. 1, a unique optimized set of values for $A_0$, $d_{max}$ and $N_{Min}$ was determined, which provided the best agreement between OKMC simulation and experiment (also shown on Tab. 2), for a given imposed Fe content in the clusters. The horizontal bars are the experimental errors reported by the authors of the studies. The vertical bars represent the range of variation of our predicted values when $0.4 \leq P_{APT} \leq 0.6$ and $0.5nm \leq M_{APT} \leq 1.5nm$. The R² values are Pearson correlation coefficients. Additional figures are provided in the supplementary materials, including the figure here without error bars, and a reproduction of the figures with one set of data at a time, for clarity.

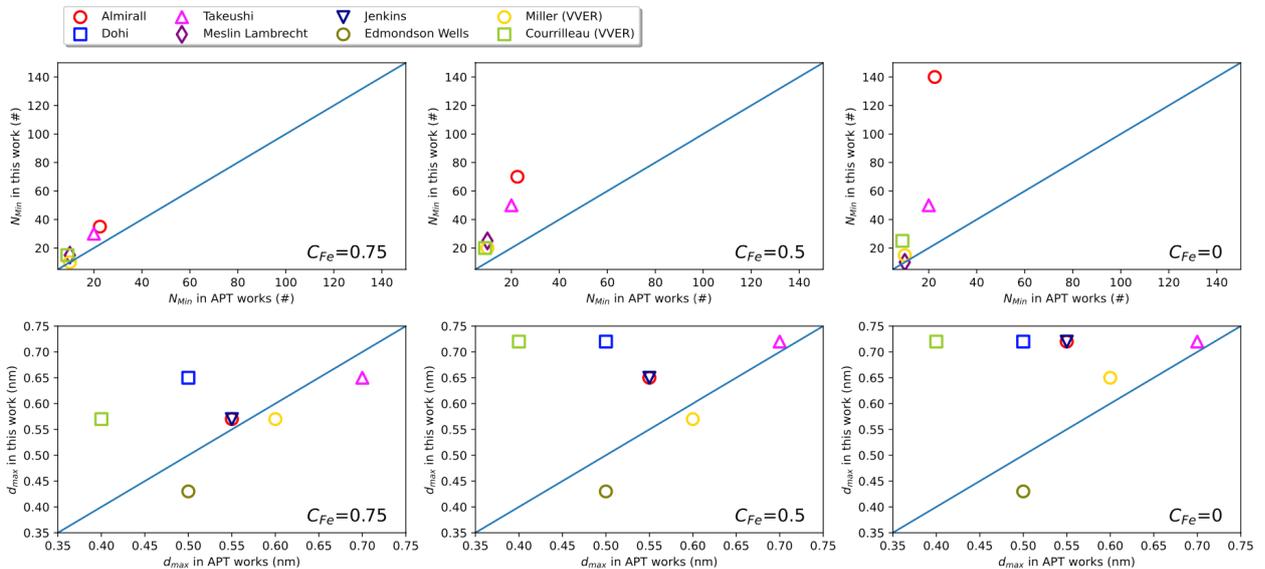

**Fig. 7** – Comparison between the $N_{Min}$ and $d_{max}$ parameters for clusters analysis reported by the APT experimental works (references in Tab. 2), and those derived in this work. The other fitted parameter, $A_0$, does not have an experimental counterpart to be directly compared with.

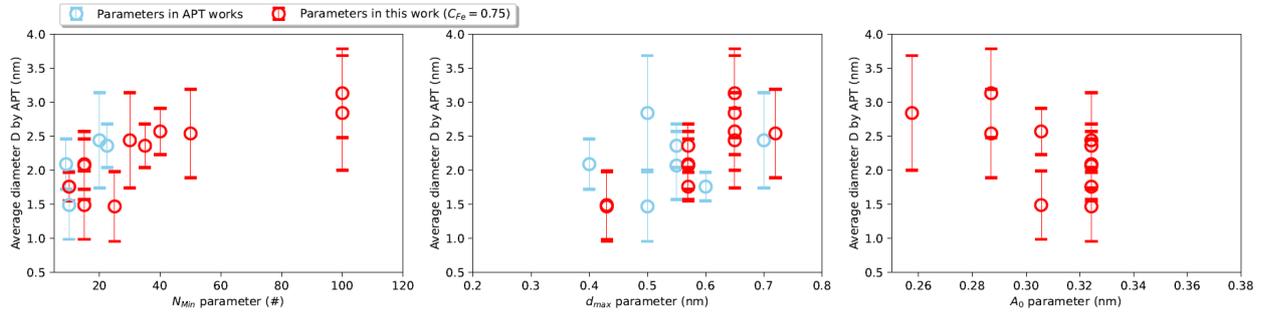

**Fig. 8** – Relation between the average diameter $D$ as reported in the APT experimental works (used as target for fitting in this work), and the $N_{Min}$, $d_{max}$, and $A_0$ parameters. For $N_{Min}$ and $d_{max}$, the figures include both the parameters used in the experimental works (Tab. 2), when reported, and those fitted in this work ($C_{Fe}$ = 0.75).

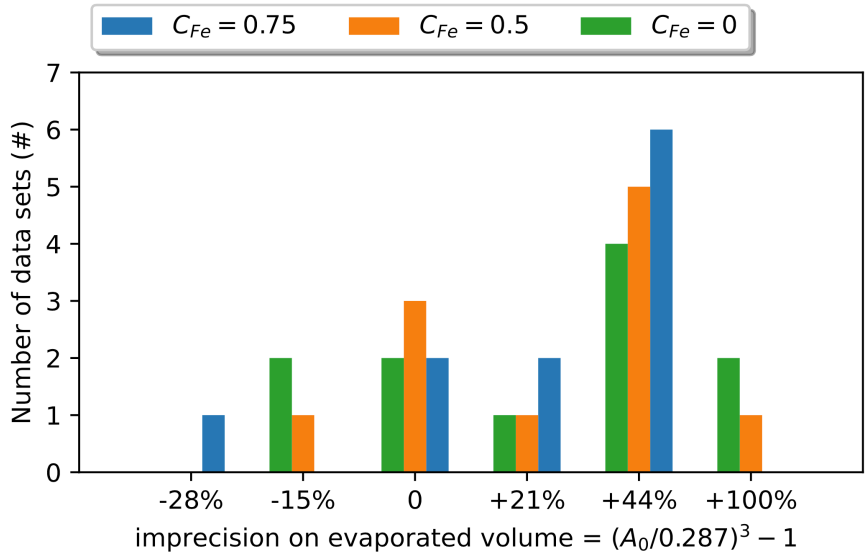

**Fig. 9** – Optimal values for the $A_0$ parameter determined in this work. For clarity, they were converted into the uncertainty of estimation for the evaporated sample volume.

## 4 Discussion

Fig. 6 and Tab. 2 provide key answers to the main question investigated in this work, i.e., up to what extent the reported number density $N$ density and average diameter $D$ of clusters derived from APT experiments are representative of the actual cluster population in RPV materials. The good agreement between our predictions and the experimental measurements shown in Fig. 6 is only possible if, as we anticipated in section 1, the different data sets are indeed differentiated by the parameters of the analysis (Tab. 2). This demonstrates that, overall, $N$ and $D$ are intrinsically imprecise descriptors that can only be directly compared if they come from consistently analysed specimens. To further demonstrate this in a visual way, we did the exercise depicted in Fig. 10. In the top panels in Fig. 10, we reproduced Fig. 1, but instead of plotting the values of *N* and *D* reported experimentally, we plotted those that are predicted by our model, using the best-match parameters determined in section 3 (Tab. 2, $C_{Fe}$ = 0.75). We see that these top panels in Fig. 10 are qualitatively similar to Fig. 1: N increases slightly with the received neutron dose (with large scatter because of the effect of chemical composition), while *D* seems to be only a function of the parameters of the cluster analysis from which it was derived. However, we can take a step forward and make all data consistent with each other, by analysing all data in the same way. It is as if we were simulating the results that would be obtained if all material samples from 11 studies were given to a 12th laboratory to make its own analysis, with its own choice of analysis parameters, consistently used for all samples. Specifically, we chose $P_{APT}$ = 0.5, $M_{APT}$ = 1.5nm, $A_0$ = 0.287 nm, $d_{max}$ = 0.43 nm and $N_{Min}$ = 15 (but a similar conclusion is reached with any other set of parameters). We show in the bottom panels in Fig. 10 the values for *N* and *D* obtained if the parameters for the clusters analysis are homogenized for all cases. We see, in Fig. 10d that the values of D collapse onto a single curve, independently of the chemical composition of the materials and of the dose rate, only very slightly increasing with received neutron dose. This substantiates our surmise that the *D* values were mainly a function of how the analysis of the APT data was performed. On the other hand, in Fig. 10c, the number densities *N* shows a more marked scaling with neutron dose, but, due to the scatter, clearly remains dependent of the chemical composition and perhaps the dose-rate, as well.

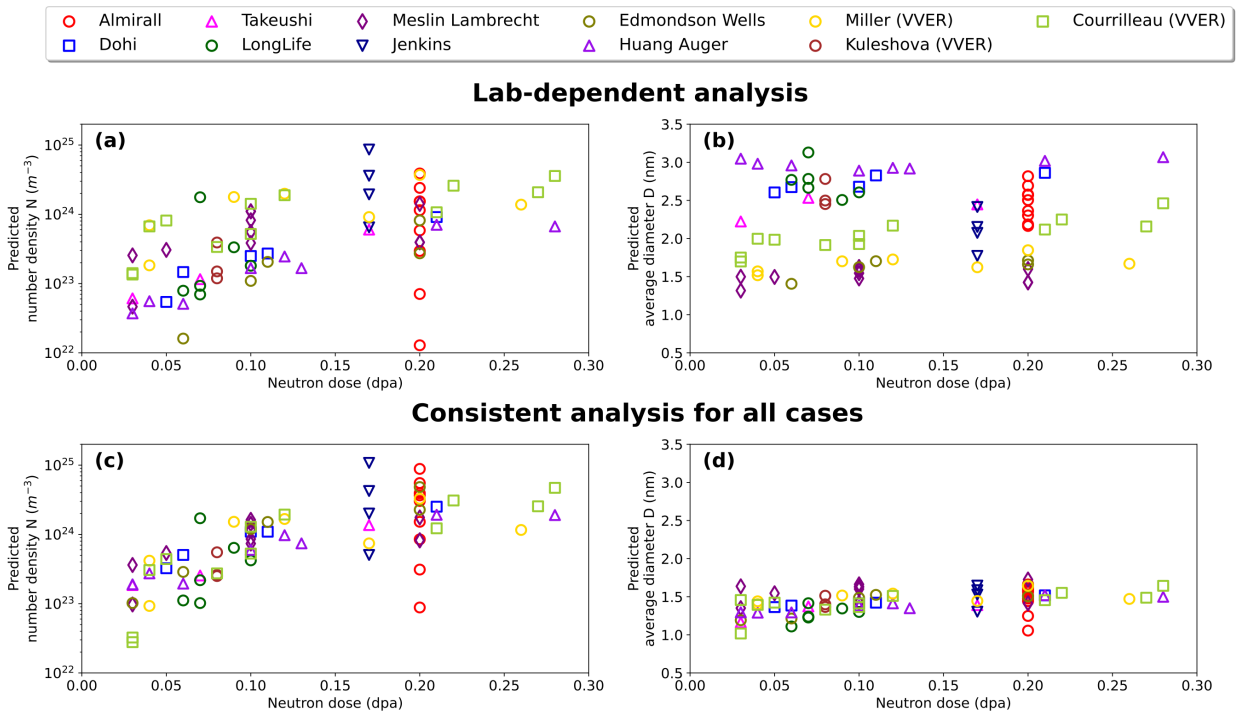

**Fig. 10** – Comparison of our model predictions when; (a, b) The best-match $N_{Min}$, $d_{max}$ and $A_0$ parameters are applied individually for each data set, as listed in Tab. 2 for the case $C_{Fe}$ = 0.75. This figure is similar to Fig. 1, except that the values of *N* and *D* are here those predicted by our model, and not those measured experimentally; (c, d) The values $N_{Min}$ = 15, $d_{max}$ = 0.43nm, $A_0$ = 0.287 and $C_{Fe}$ = 0.75 are consistently applied to all cases, regardless of in which experimental set they belong.

An interesting result that Fig. 6 highlights is that $C_{Fe}$ = 0.75 provides the most realistic predictions of *N* and *D*, from all standpoints. Indeed, $C_{Fe}$ = 0.75 resulted in the lowest residual discrepancy between our predicted values and those reported experimentally (see Tab. 2), but also the best agreement with the $N_{Min}$ and $d_{max}$ parameters that were used in the analysis of the experimental data (Fig. 7), as well as the most realistic range for the $A_0$ parameter (Fig. 9). The original purpose of the $C_{Fe}$ parameter as defined in section 2.2 is to explicitly account for the uncertainty regarding the true Fe content of solute clusters, i.e., before the measurement is performed. However, as explained in section 2.3, our model does not include trajectory aberration effects, which may also increase the apparent Fe content of solute clusters in experimentally reconstructed atomic structures. The $C_{Fe}$ parameter as defined in our model thus unavoidably accounts for both effects, which cannot be dissociated in this study. Therefore, we cannot conclude that the true Fe content in solute clusters is large. We can only note that our model delivers the most realistic predictions when $C_{Fe}$ = 0.75, which is consistent with the experimental observations. Whichever the reason, the results of Tab. 2 can be considered as correct trends with respect to the $C_{Fe}$ parameter.

It is noteworthy that one outlier systematically remains in the lower panels in Fig. 6: for that specific point the model unavoidably underestimates $D$ compared to the experimentally reported value. This is in fact the one case for which Huang and Auger reported D = 4nm, while they reported D = 3 nm for all others. The prediction by our model if $C_{Fe}$ = 0.75 is 3 nm also for that point.

Another notable aspect in Fig. 6 is that we assumed that each laboratory processed their reconstructed APT data, collected from the measurements, in a consistent way for all samples, regardless of the source material and regardless of their irradiation conditions. Accordingly, we applied a methodology based on a systematic exploration of all combinations of parameters and based on a minimization of residual discrepancies for the whole dataset. However:

- When measured APT 3D configuration are analysed, each researcher is likely to adapt the parameters of the analysis, i.e., $N_{Min}$ and $d_{max}$, as well as others, for every single measurement, in an iterative manner, depending on the focus of the study, and also depending on the results that are being obtained. For example, we anticipate that analyses which reveal low number densities of clusters are likely parameterized differently than those revealing large number densities, e.g. using larger $d_{max}$ and lower $N_{min}$ values. Our methodology cannot catch all the subtleties within each of the 11 data sets defined in Tab. 1 and Tab. 2.

- Even if $N_{Min}$ and $d_{max}$ were in fact homogenized for all analyses within a data set, the uncertainties on the evaporated volumes of each individual samples, which we account for via the $A_0$ parameter, would remain. It is key to bear in mind that this uncertainty is a priory unique for each individual measurement, independently of the others. Our assumption in this work that a unique value of $A_0$ can be used to simulate the analysis performed by each lab was instrumental, for the sake of practicality, and to avoid introducing an excess of free parameters in the fit performed via Eq. (3-5). Notwithstanding, the fact that a satisfactory agreement could be obtained in Fig. 6 suggests that there is indeed a trend for the $A_0$ parameters to be similar for all measurements performed by a given lab. The intrinsic and unavoidable fluctuations of this parameter between individual measurements is reflected by the fact that, eventually, our predictions of the number densities $N$ tend to be somewhat less accurate than those of $D$, in terms of correlation parameter.

Having demonstrated that $N$ and $D$ are intrinsically imprecise descriptors, that mask physical trends with analysis dependent trends, how can we therefore compare findings reported by different labs? Our results in section 3 provide valuable clues in this respect. To start, we predicted that the parameters $P_{APT}$ and $M_{APT}$ have a secondary impact on the derived values of $N$ and $D$; therefore, the characteristics of the APT setup itself should not be at issue. In contrast, as expected, the choice of the $d_{max}$ and $N_{min}$ parameters have a strong influence on the derived $N$ and $D$. Fortunately, these parameters are deliberately chosen during the analysis, in a fully controlled way; it is thus always possible to perform several cluster analyses out of a given measurement, with varying parameters, which would allow easier comparisons with the findings of other laboratories. There is, however, one parameter in this work, $A_0$, that is not directly under control during an experimental APT measurement. Indeed, we introduced this parameter to account for the inaccuracy of estimation of the evaporated sample volume, which is realistically anticipated to be as high as 50% in most cases. Consequently, even in the ideal situation where two sets of data were purposefully processed using identical clusters 'analysis methods, with identical $d_{max}$ and $N_{min}$ parameters, the uncertainties on the evaporated volumes of each sample will always remain. The consequences of these uncertainties will mostly affect the number density $N$, as illustrated in Fig. 11. This figure collects only the data sets for which the best-match $A_0$ was 0.3243 nm, with $C_{Fe} = 0.75$, and compare the derived $N$ and $D$ with those derived when $A_0$ is changed to 0.287 nm. This exercise mimics an attempt of APT findings comparison, where a measurement is processed using $N_{min}$ and $d_{max}$ values that correspond to the analysis performed by another lab, but there is a discrepancy in the accuracies of evaporated volume estimation which we simulate by deliberately changing the value of $A_0$. Comparing both left and right panels in Fig. 11, we see that decreasing $A_0$ has mainly two effects:

1. The distance between all pairs of atoms in the reconstructed sample is decreased. Because we assumed in Fig. 11 that $N_{Min}$ and $d_{max}$ are constant, the effects are:

   1.1. Many small solute clusters (constituted in the source configuration of nearly $N_{Min}$ / $P_{APT}$ atoms) that were not detected with a larger $A_0$ are now detected, because their atoms are closer to each other and satisfy the conditions to be identified as clusters given the constant $d_{max}$.

   1.2. A priori all solute clusters that were already detected grow in size, because more atoms further away from the centre of mass of the clusters are reached by the constant $d_{max}$.

2. The volume of the reconstructed sample is decreased.

On the one hand, both the effects 1.1 and 2 listed above contribute to increasing the number density $N$. The effect 1.2 may in fact decrease $N$ in case of direct coalescence of clusters (see Fig. 5), but it is insignificant in Fig. 11. On the other hand, the effect 1.1 tends to decrease the average diameter $D$, because new clusters with a small diameter are added, while the effect 1.2 tends to increase $D$. This competition of opposite effects explains why $D$ is found to be almost unaffected in Fig. 11. This finding is an aspect that is probably not emphasized enough when APT laboratories compare their results, even when identical cluster analysis methods are employed. These effects can be minimized by application of formulas which avoid spatial coordinates from the APT but use a mix of relative (a ratio of clustered to all ions detected) and theoretical (matrix density) components.

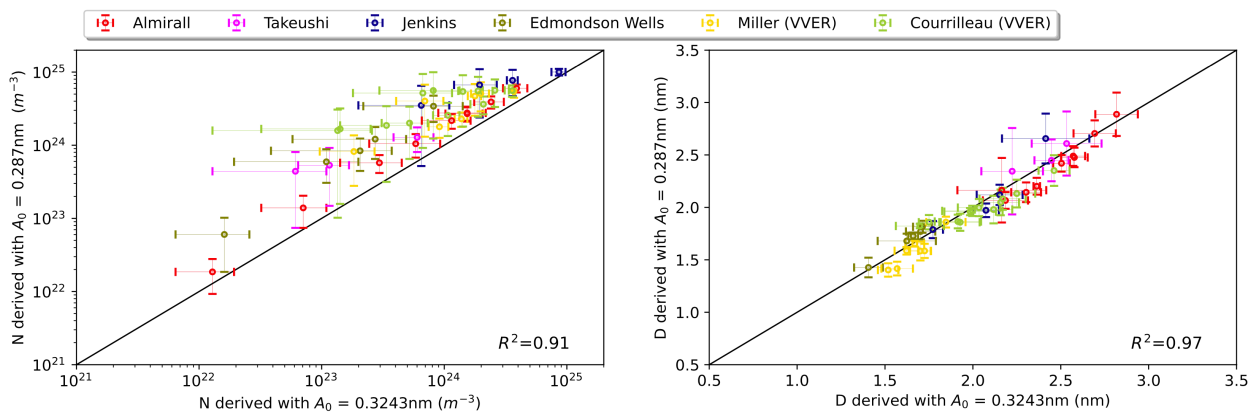

**Fig. 11** – Comparison of our model predictions when different values of A₀ are used. The values of $N_{min}$ and $d_{max}$ are the best-match listed in Tab. 2, with $C_{Fe}$ = 0.75. The horizontal bars are the experimental errors reported by the authors of the studies. The vertical bars represent the range of variation of our predicted values when $0.4 \leq P_{APT} \leq 0.6$ and $0.5nm \leq M_{APT} \leq 1.5nm$. The R² values are Pearson correlation coefficients.

Unsurprisingly, the most influential parameter is found to be $N_{Min}$. This parameter defines the cluster size (in terms of number of detected ions) below which a detected cluster is considered as noise. Since the reconstructed atomic structures (Fig. 2cd) are largely distorted from the true material matrix ($M_{APT} \gg A_0$), retaining the smallest clusters from the analysis (e.g., choosing $N_{Min} \leq 5$) and being confident that they indeed correspond to a real cluster in the source sample is not straightforward. Decreasing $N_{Min}$ increases the risk to retain *false positives*, which are clusters of solute atoms formed by random statistical fluctuations (see, e.g., Ref. [63]), even for idealized configurations as constructed in this work (Fig. 3), where clusters are assumed to be spheres before evaporation. As we see in Fig. 5, given $N_{Min}$, the smallest clusters ($D$ < 1.5nm) are detected with a probability that gradually increases from 0 to 1 as a function of increasing cluster size (Fig. 2e). This stochasticity originates from the fact that all solute clusters contain a significant concentration of Fe, which is explicitly introduced from the reference configuration via the $C_{Fe}$ parameter (Fig. 3), but also implicitly added during the measurement due to the application of lattice distortions ($M_{APT}$ parameter). The solute clusters in the atomic configuration constructed by our modelled APT analysis are hence always found as random dilute alloys, even if $C_{Fe}$ = 0 from our reference configurations. Thus, solute atoms are positioned in random locations around the centre of mass of their original clusters (see Fig. 3 and Fig. 5cdef). Consequently, some of the smallest solute clusters originally present in the reference configuration are not identified as cluster after the APT measurement because their solute atoms (those that are detected, given $P_{APT}$) are reconstructed at too far apart positions. We have checked that all the clusters in the reference configuration are fully recovered by the APT measurement only if $N_{Min}$ = 5, $P_{APT} \approx 1$ and $M_{APT} \approx 0$. But in practice, for real APT measurements of real samples, this set of parameters is not realistic ($P_{APT} < 1$ and $M_{APT} > 0$) and $N_{Min} \geq 10$ is a common choice (see Tab. 2). An immediate consequence is that there is potentially a large number density ($\geq 10^{24}$ m⁻³) of small solute clusters (d ≤ 1.5nm) that are systematically unresolved in APT (Fig. 5). This fact needs to be accounted for when comparing simulation results with experimental findings, as counts made in simulations will include clusters of all sizes, down to the smallest ones, so one needs to know which ones should be excluded for the comparison to be meaningful. Moreover, on a case-by-case basis, this neglected population of clusters might have an influence on hardening and embrittlement that an assessment from experimental values might hide or distort. The model proposed in [5] provides an estimation of the radiation-induced hardening (in MPa) as $\Delta\sigma_y \sim 1930\sqrt{ND^3}$. For example, the solute clusters unresolved by APT in Fig. 5 roughly correspond to $N$ = 2×10²⁴ m⁻³ and $D$ = 0.75nm, for which the model predicts $\Delta\sigma_y$ = 40 MPa, which is not negligible compared to the experimental hardening value 291 MPa that was reported by Almirall et al. in [30]. That is, the APT-invisible cluster population might have an influence on the hardening and embrittlement. The model we developed and used, which produces realistic microstructures for RPV steels of given composition, irradiated under specific conditions, has, in this respect, a high added value, as it can be used to complement APT experiments. Firstly, it provides a reference for what seems reasonable to expect from the APT study. Secondly, it gives an assessment of the cluster population that APT cannot see, which may help understand the origin of possible departures from the $\Delta\sigma_y \sim 1930\sqrt{ND^3}$ law. Finally, by completing the information with the presence or not of vacancies or self-interstitials in the clusters, it may help connect APT results with results from other microstructural examination techniques, such as positron annihilation (sensitive to the presence of vacancies) and electron microscopy (sensitive to the presence of self-interstitial clusters).

# 5 Conclusions

In this work we developed a methodology that enabled us to study how the characterization by APT studies of solute clusters in RPV materials is influenced by the intrinsic parameters of the technique, and the parameters of the subsequent data processing. Importantly, we made a distinction between the parameters that are directly under the control of the operators ($N_{Min}$ and $d_{max}$ in this study), and those that are not ($P_{APT}$, $M_{APT}$ and $A_0$). Our methodology is based on the use of a physical model for microstructure evolution in RPV steels under irradiation, which produces plausible reference microstructures that are confidently representative of the real microstructure before the APT measurement is performed. This enables the prediction of the population of solute clusters that is not detected during the cluster's identification phase, as a function of the parameters of the analysis (see Fig. 5). Accordingly, we showed that when the findings of APT studies are summarized with the key descriptors $N$ (number density) and $D$ (diameter), valuable information about the true clusters population is lost, and significantly different results are likely to be derived by different labs, given the same samples. These descriptors are hence intrinsically imprecise, as demonstrated in Fig. 6 and Fig. 10. Our results in Fig. 5 also highlight that this intrinsic imprecision is mainly due to true distribution of cluster sizes (before the measurement is performed), which are dominantly small ($D < 2$nm). Consequently, trends with important parameters such as neutron dose, dose-rate and temperature, will be masked when $N$ and $D$ data from different laboratories are combined.

A logical recommendation stemming from this work is that APT studies of RPV materials should not be summarized into single values of N and D. Ideally, detailed histograms of cluster sizes distributions should be provided, with varying $N_{Min}$ and $d_{max}$ parameters, which would allow more accurate comparisons of the findings between different labs. However, we highlighted that there remains the uncertainty on the evaporated volume (through the $A_0$ parameter), which is probably an aspect that has not been emphasized enough by the community.

The results in this work also highlight that physical models for neutron irradiation in steels, such as our OKMC model employed here, provide valuable information that can help for the analysis and interpretation of APT findings. Such models enable the fast production of simulated microstructures that correspond to the cases analysed by APT in various laboratories. These microstructures, after verification that they are indeed consistent with the experiment (via best match analysis parameters), can subsequently be used to remove the interlaboratory experimental scatter, in order to reveal the actual trends of the data ($N$, $D$) with the important variables (neutron dose, dose-rate, temperature …). For example, our model showed that $D$ only increases very slightly with the received neutron dose, independently of steel composition and probably also received dose-rate, while the number density $N$ does increase with dose, but remains also significantly dependent on composition and, perhaps, dose-rate, as well. Incidentally, these predictions advocate for a radiation-induced process for solute clustering in RPV steels, rather than for a nucleation and growth mechanism, that should show $D$ increase (growth) and $N$ decrease (coarsening) above a certain dose. Our model also provides a tool to assess what the APT analysis cannot see, thereby complementing the experimental technique: small clusters, presence of point-defects in solute clusters, etc. This facilitates a comparison with complementary techniques that are sensitive to point-defect clusters, as well as the identification of possible departures from known trends when the hardening and embrittlement is assessed from microstructural considerations.

# Acknowledgements

This work received partial financial support in the framework of the 2019-2020 Euratom research and training programme (ENTENTE project, grant No 900018) and contributes to the Joint Programme on Nuclear Materials of the European Energy Research Alliance (EERA JPNM, IOANIS2 pilot project). The views and opinions expressed herein do not necessarily reflect those of the European Commission.

# CRediT author statement

**N. Castin and P. Klupś:** Conceptualization, Methodology, Formal analysis, Visualization, Software. **G. Bonny and M.I. Pascuet:** Formal analysis, Visualization. **M. Konstantinovic, M. Moody and L. Malerba:** Conceptualization, Supervision. **All:** Writing - Original Draft, Writing - Review & Editing.

# Data availability

The raw/processed data required to reproduce these findings cannot be shared at this time as the data also forms part of an ongoing study.

# References

[1]     M.K. Miller, K.F. Russell, Embrittlement of RPV steels: An atom probe tomography perspective, Journal of Nuclear Materials 371 (2007) 145-160.

[2]     G.R. Odette, T. Yamamoto, T. J. Williams, R. K. Nanstad, and C. A. English, On the history and status of reactor pressure vessel steel ductile to brittle transition temperature shift prediction models, *Journal of Nuclear Materials* 526 (2019) 151863.

[3]     Y. Hashimoto, A. Nomoto, M. Kirk, K. Nishida, Development of new embrittlement trend curve based on Japanese surveillance and atom probe tomography data, Journal of Nuclear Materials 553 (2021) 153007.

[4]     N. Castin et al. The dominant mechanisms for the formation of solute-rich clusters in low-Cu steels under irradiation, Mater. Today Energy 17 (2020) 100472.

[5]     N. Castin et al., Multiscale modelling in nuclear ferritic steels: From nano-sized defects to embrittlement, Materials Today Physics 27 (2022) 100802.

[6]     E.A. Marquis, M. Bachhav, Y. Chen, Y. Dong , L.M. Gordon, A. McFarland, On the current role of atom probe tomography in materials characterization and materials science, Current Opinion in Solid State and Materials Science 17 (2013) 217–223.

[7]     E.A. Marquis at al., Nuclear reactor materials at the atomic scale, Materials Today 12 (2009) 30-37.

[8]     Wang, X., Hatzoglou, C., Sneed, B. *et al.,* Interpreting nanovoids in atom probe tomography data for accurate local compositional measurements. *Nat Commun* **11**, 1022 (2020).

[9]     Y. Yu et al., Revealing nano-chemistry at lattice defects in thermoelectric materials using atom probe tomography, Materials Today 32 (2020) 260-274.

[10]    M.K. Miller, T.F. Kelly, K. Rajan, S.P. Ringer, The future of atom probe tomography, Materials Today 15 (2012) 158-165.

[11]    P. D. Edmondson, C. M. Parish, and R. K. Nanstad, Using complimentary microscopy methods to examine Ni-Mn-Si-precipitates in highly-irradiated reactor pressure vessel steels, Acta Materialia 134 (2017) 31–39.

[12]    B. Gómez-Ferrer, C. Dethloff, E. Gaganidze, L. Malerba, C. Hatzoglou, C. Pareige, Nano-hardening features in high-dose neutron irradiated Eurofer97 revealed by atom-probe tomography, Journal of Nuclear Materials 537 (2020) 152228.

[13]    T.G. Lach, W.E. Frazier, J. Wang, A. Devaraj, T.S. Byun, Precipitation-site competition in duplex stainless steels: Cu clusters vs spinodal decomposition interfaces as nucleation sites during thermal aging, Acta Materialia 196 (2020) 456-469.

[14]    Q. Yang et al., Cu precipitation in electron-irradiated iron alloys for spent-fuel canisters, Journal of Nuclear Materials 572 (2022) 154038.

[15]    J.P. Balbuena, L. Malerba, N. Castin, G. Bonny, M.J. Caturla, An object kinetic Monte Carlo method to model precipitation and segregation in alloys under irradiation, Journal of Nuclear Materials 557 (2021) 153236.

[16]    W.E. Frazier, T.G. Lach, T.S. Byun, Monte Carlo simulations of Cu/Ni-Si-Mn co-precipitation in duplex stainless steels, Acta Materialia 194 (2020) 1-12.

[17]    J. Emo, C. Pareige, S. Saillet, C. Domain, P. Pareige, Kinetics of secondary phase precipitation during spinodal decomposition in duplex stainless steels: A kinetic Monte Carlo model – Comparison with atom probe tomography experiments, journal of nuclear materials 451 (2014) 361-365.

[18]    Z. Mao, C. Booth-Morrison, C.K. Sudbrack, G. Martin, D.N. Seidman, Kinetic pathways for phase separation: An atomic-scale study in Ni–Al–Cr alloys, Acta Materialia 60 (2012) 1871–1888.

[19]    N. Castin, L. Messina, C. Domain, R. C. Pasianot, and P. Olsson, Improved atomistic Monte Carlo models based on ab-initio-trained neural networks: Application to FeCu and FeCr alloys, Phys. Rev. B 95 (2017) 214117.

[20]    N. Castin and L. Malerba, Calculation of proper energy barriers for atomistic kinetic Monte Carlo simulations on rigid lattice with chemical and strain field long-range, effects using artificial neural networks, the journal of chemical physics 132 (2010) 074507.

[21]    N. Castin, M.I. Pascuet and L. Malerba, Modeling the first stages of Cu precipitation in α-Fe using a hybrid atomistic kinetic Monte Carlo approach, the journal of chemical physics 135 (2011) 064502.

[22]    Moody, M., Ceguerra, A., Breen, A. *et al.* Atomically resolved tomography to directly inform simulations for structure–property relationships. *Nat Commun* 5 (2014) 5501.

[23]    A. Prakash et al., Atom probe informed simulations of dislocation–precipitate interactions reveal the importance of local interface curvature, Acta Materialia 92 (2015) 33-45.

[24]    G. Monnet, Multiscale modeling of irradiation hardening: application to important nuclear materials, J. Nucl. Mater. 508 (2018) 609-627

[25]    J.M. Hyde et al., Analysis of Radiation Damage in Light Water Reactors: Comparison of Cluster Analysis Methods for the Analysis of Atom Probe Data, Microscopy and Microanalysis 23 (2017) 366–375.

[26]    J.M. Hyde, E.A. Marquis, K.B. Wilford, T.J. Williams, A sensitivity analysis of the maximum separation method for the characterisation of solute clusters, Ultramicroscopy 111 (2011) 440-447.

[27]    F. De Geuser, B. Gault, Metrology of small particles and solute clusters by atom probe tomography, Acta Materialia 188 (2020) 406-415.

[28]    E.A. Marquis, V. Araullo-Peters, Y. Dong, A. Etienne, S. Fedotova, et al. (2019). On the use of density-based algorithms for the analysis of solute clustering in atom probe tomography data. In

Proceedings of the 18th International Conference on Environmental Degradation of Materials in Nuclear Power Systems–Water Reactors (pp. 2097-2113). Springer International Publishing

[29]   Y. Dong et al., Atom probe tomography interlaboratory study on clustering analysis in experimental data using the maximum separation distance approach, Microscopy and Microanalysis 25 (2019) 356-366.

[30]   N. Almirall et al. Precipitation and hardening in irradiated low alloy steels with a wide range of Ni and Mn compositions. Acta Mater. 179, 119–128 (2019).

[31]   P. Auger, P. Pareige, S. Welzel, and J.-C. Van Duysen, Synthesis of atom probe experiments on irradiation-induced solute segregation in French ferritic pressure vessel steels, Journal of Nuclear Materials 280 (2000) 331–344.

[32]   H. Huang, B. Radiguet, P. Todeschini, G. Chas, and P. Pareige, Atom Probe Tomography characterization of the microstructural evolution of a low copper reactor pressure vessel steel under neutron irradiation, MRS Proc. 1264 (2010) 1264-BB05-18.

[33]   B. M. Jenkins et al., The effect of composition variations on the response of steels subjected to high fluence neutron irradiation, Materialia 11 (2020) 100717.

[34]   M. Lambrecht et al., On the correlation between irradiation-induced microstructural features and the hardening of reactor pressure vessel steels, Journal of Nuclear Materials 406 (2010) 84–89.

[35]   E. Meslin et al., Characterization of neutron-irradiated ferritic model alloys and a RPV steel from combined APT, SANS, TEM and PAS analyses, Journal of Nuclear Materials 406 (2010) 73–83.

[36]   E. Meslin, B. Radiguet, P. Pareige, and A. Barbu, Kinetic of solute clustering in neutron irradiated ferritic model alloys and a French pressure vessel steel investigated by atom probe tomography, Journal of Nuclear Materials 399 (2010) 137–145.

[37]   E. Meslin, B. Radiguet, P. Pareige, C. Toffolon, and A. Barbu, Irradiation-Induced Solute Clustering in a Low Nickel FeMnNi Ferritic Alloy, Exp Mech 51 (2011) 1453–1458.

[38]   P.D. Edmondson, M.K. Miller, K.A. Powers, R.K. Nanstad, Atom probe tomography characterization of neutron irradiated surveillance samples from the R. E. Ginna reactor pressure vessel, Journal of Nuclear Materials 470 (2016) 147-154.

[39]   P. Wells, The Character, Stability and Consequences of Mn-Ni-Si Precipitates in Irradiated Reactor Pressure Vessel Steels, PhD thesis dissertation, 2016, Santa Barbara, https://escholarship.org/uc/item/3vh4k9tf.

[40]   C. Courilleau, B. Radiguet, R. Chaouadi, E. Stergar, A. Duplessi, P. Pareige, Contributions of Ni-content and irradiation temperature to the kinetic of solute cluster formation and consequences on the hardening of VVER materials, Journal of Nuclear Materials 585 (2023) 154616.

[41]   M.K. Miller et al. Evolution of the nanostructure of VVER-1000 RPV materials under neutron irradiation and post irradiation annealing. J. Nucl. Mater. 385 (2009) 615–622.

[42]   E. A. Kuleshova et al., Specific Features of Structural-Phase State and Properties of Reactor Pressure Vessel Steel at Elevated Irradiation Temperature, Science and Technology of Nuclear Installations 2017 (2017) 1–12.

[43]   E. A. Kuleshova et al., Study of the flux effect nature for VVER-1000 RPV welds with high nickel content, Journal of Nuclear Materials 483 (2017) 1–12.

[44]   E. A. Kuleshova et al., Mechanisms of radiation embrittlement of VVER-1000 RPV steel at irradiation temperatures of (50–400)°C, Journal of Nuclear Materials 490 (2017) 247–259.

[45]   K. Dohi, K. Nishida, A. Nomoto, N. Soneda, H. Matsuzawa, and M. Tomimatsu, Effect of Neutron Flux at High Fluence on Microstructural and Hardness Changes of RPV Steels, in ASME 2010 Pressure Vessels and Piping Conference: Volume 9, Bellevue, Washington, USA, Jan. 2010, pp. 95–102, doi: 10.1115/PVP2010-25514.

[46]   T. Takeuchi et al., Effects of chemical composition and dose on microstructure evolution and hardening of neutron-irradiated reactor pressure vessel steels, Journal of Nuclear Materials 402 (2010) 93–101.

[47]    L.T. Stephenson, M.P. Moody, P.V. Liddicoat, S.P. Ringer, New Techniques for the Analysis of Fine-Scaled Clustering Phenomena within Atom Probe Tomography (APT) Data, Microscopy and Microanalysis 13 (2007) 448-463.

[48]    M. Chiapetto, L. Malerba, C.S. Becquart, Effect of Cr content on the nanostructural evolution of irradiated ferritic/martensitic alloys: An object kinetic Monte Carlo model, Journal of Nuclear Materials 465 (2015) 326-336.

[49]    L. Malerba, C.S. Becquart, C. Domain, Object kinetic Monte Carlo study of sink strengths, Journal of Nuclear Materials 360 (2007) 159-169.

[50]    B. Gómez-Ferrer, C. Dethloff, E. Gaganidze, L. Malerba, C. Hatzoglou, C. Pareige, Nano-hardening features in high-dose neutron irradiated Eurofer97 revealed by atom-probe tomography, Journal of Nuclear Materials 537 (2020) 152228.

[51]    E. Altstadt et al., NUGENIA position on RPV Irradiation Embrittlement issues based on the outcome of the EURATOM FP7 project LONGLIFE - RPV Irradiation Embrittlement (2015). https://snetp.eu/wp-content/uploads/2020/06/NUGENIA_position_paper_RPV_irradiation_embrittlement_May_2015.pdf;

[52]    Ulbricht A., Dykas J., Chekhonin P., Altstadt E., Bergner F., Small-angle neutron scattering study of neutron-irradiated and post-irradiation annealed VVER-1000 reactor pressure vessel weld material, Frontiers in Nuclear Engineering 2 (2023). https://www.frontiersin.org/articles/10.3389/fnuen.2023.1176288

[53]    Shu, S., Wirth, B. D., Wells, P. B., Morgan, D. D. & Odette, G. R. Multi-technique characterization of the precipitates in thermally aged and neutron irradiated Fe-Cu and Fe-Cu-Mn model alloys: Atom probe tomography reconstruction implications. Acta Mater. 146, 237-252 (2018).

[54]    J.M. Hyde, M.G. Burke, G.D.W. Smith, P. Styman, H. Swan, K. Wilford, Uncertainties and assumptions associated with APT and SANS characterisation of irradiation damage in RPV Steels, Journal of Nuclear Materials 449 (2014) 308–314.

[55]    M.K. Miller, B.D. Wirth, G.R. Odette, Precipitation in neutron-irradiated Fe-Cu and Fe-Cu-Mn model alloys: a comparison of APT and SANS data, Materials Science and Engineering A353 (2003) 133-139.

[56]    F. Bergner, C. Pareige, V. Kuksenko, L. Malerba, P. Pareige, A. Ulbricht, A. Wagner, Critical assessment of Cr-rich precipitates in neutron-irradiated Fe–12 at%Cr: Comparison of SANS and APT, Journal of Nuclear Materials 442 (2013) 463–469.

[57]    S. Mühlbauer, D. Honecker, E.A. Périgo, F. Bergner, S. Disch, A. Heinemann, S. Erokhin, D. Berkov, C. Leighton, M.R. Eskildsen, A. Michels, Magnetic small-angle neutron scattering, Reviews of Modern Physics 91 (2019) 015004.

[58]    F. Vurpillot, C. Oberdorfer, Modeling Atom Probe Tomography: A review, Ultramicroscopy 159 (2015) 202-216.

[59]    M.P. Moody, B. Gault, L.T. Stephenson, D. Haley, S.P. Ringer, Qualification of the tomographic reconstruction in atom probe by advanced spatial distribution map techniques, Ultramicroscopy 109 (2009) 815–824.

[60]    T.J. Prosa, D. Olson, B. Geiser, D.J. Larson, K. Henry, and E. Steel, Analysis of implanted silicon dopant profiles, Ultramicroscopy 132 (2013) 179-185.

[61]    B.M. Jenkins, A. J. London, N. Riddle, J.M. Hyde, P.A. Bagot, M.P. Moody, Using alpha hulls to automatically and reproducibly detect edge clusters in atom probe tomography datasets, Materials Characterization 160 (2020) 110078.

[62]    P.D. Styman, J.M. Hyde, K. Wilford, G.D.W. Smith, Quantitative methods for the APT analysis of thermally aged RPV steels, Ultramicroscopy **132** (2013) 258-264.

[63]    A. Cerezo, L. Davin, Aspects of the observation of clusters in the 3-dimensional atom probe, Surf. Interface Anal. 39 (2007) 184-188.

## Supplementary Information

**Fig. SI1 –** (Absence of) correlations between the reported $N$ and $D$ and the irradiation conditions for the materials included in the various data sets defined in Tab. 1 (main manuscript). Only a very slight decrease of $D$ vs. dose rate irradiation temperature, and an increase of $N$ with dose, may perhaps be visually deduced, although with very large uncertainty.

**Fig. SI2 –** (Absence of) correlations between the reported $N$ and $D$ and the chemical contents in the materials included in the various data sets defined in Tab. 1 (main manuscript).

**Fig. SI3 –** Reproduction of Fig. 6 (main manuscript), without error bars.

**Fig. SI4 –** Reproduction of Fig. 6 (main manuscript), for $C_{Fe}$=0.75, with only the data point from the Almirall set.

**Fig. SI5 –** Reproduction of Fig. 6 (main manuscript), for $C_{Fe}$=0.75, with only the data point from the Courrillea set.

**Fig. SI6 –** Reproduction of Fig. 6 (main manuscript), for $C_{Fe}$=0.75, with only the data point from the Dohi set.

**Fig. SI7 –** Reproduction of Fig. 6 (main manuscript), for $C_{Fe}$=0.75, with only the data point from the Edmondson-Wells set.

**Fig. SI8 –** Reproduction of Fig. 6 (main manuscript), for $C_{Fe}$=0.75, with only the data point from the Huang-Auger set.

**Fig. SI9 –** Reproduction of Fig. 6 (main manuscript), for $C_{Fe}$=0.75, with only the data point from the Jenkins set.

**Fig. SI10 –** Reproduction of Fig. 6 (main manuscript), for $C_{Fe}$=0.75, with only the data point from the Kuleshova set.

**Fig. SI11 –** Reproduction of Fig. 6 (main manuscript), for $C_{Fe}$=0.75, with only the data point from the LongLife set.

**Fig. SI12 –** Reproduction of Fig. 6 (main manuscript), for $C_{Fe}$=0.75, with only the data point from the Meslin-Lambrecht set.

**Fig. SI13 –** Reproduction of Fig. 6 (main manuscript), for $C_{Fe}$=0.75, with only the data point from the Takeushi set.

**Fig. SI14 –** Reproduction of Fig. 6 (main manuscript), for $C_{Fe}$=0.75, with only the data point from the Miller set.

**Tab SI1 –** Detailed information for the materials and irradiation conditions considered in this work (Part 1/3).

**Tab SI2 –** Detailed information for the materials and irradiation conditions considered in this work (Part 2/3).

**Tab SI3 –** Detailed information for the materials and irradiation conditions considered in this work (Part 3/3).

# S1 Additional figures to Fig. 1

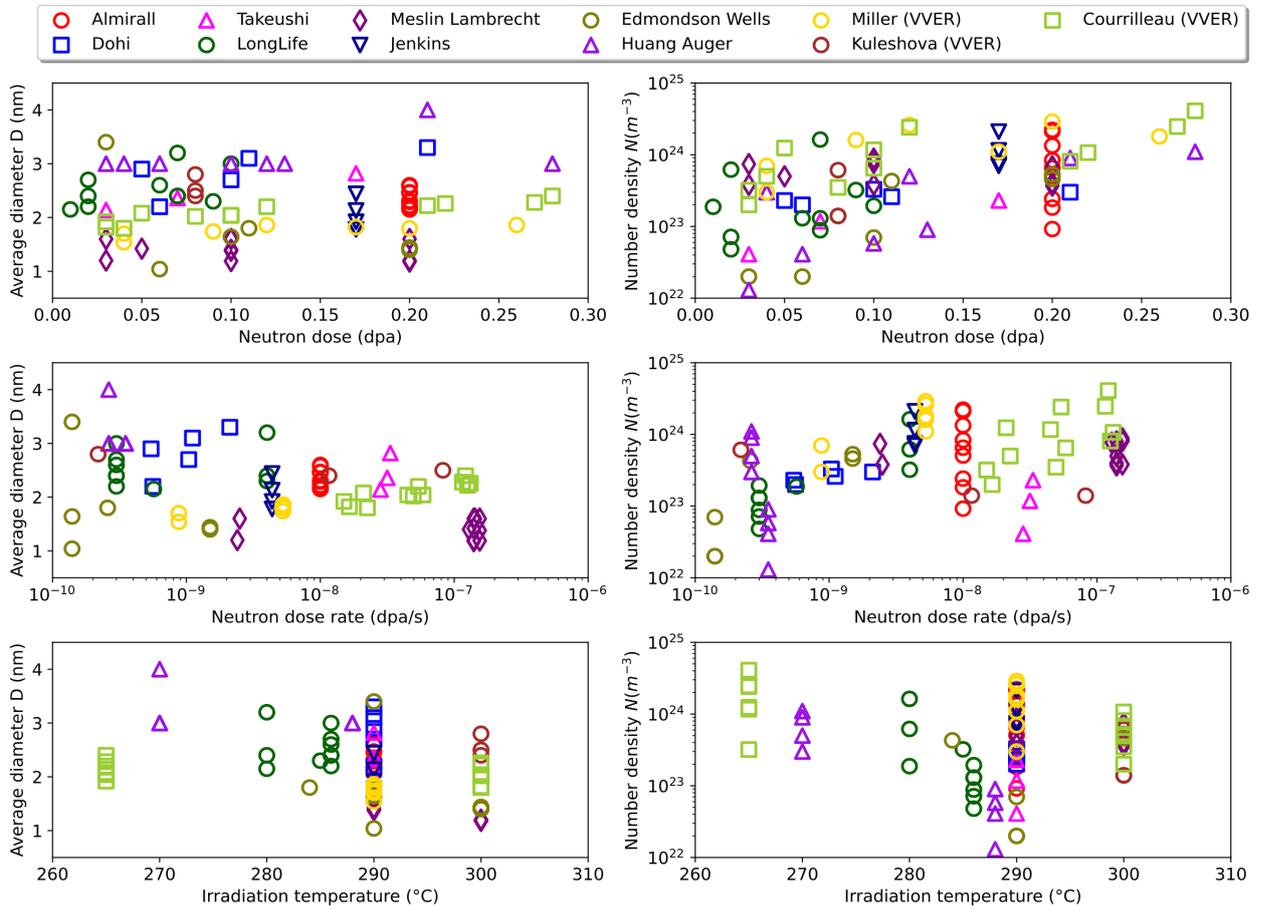

**Fig. SI1** − (Absence of) correlations between the reported $N$ and $D$ and the irradiation conditions for the materials included in the various data sets defined in Tab. 1 (main manuscript). Only a very slight decrease of $D$ vs. dose rate irradiation temperature, and an increase of $N$ with dose, may perhaps be visually deduced, although with very large uncertainty.

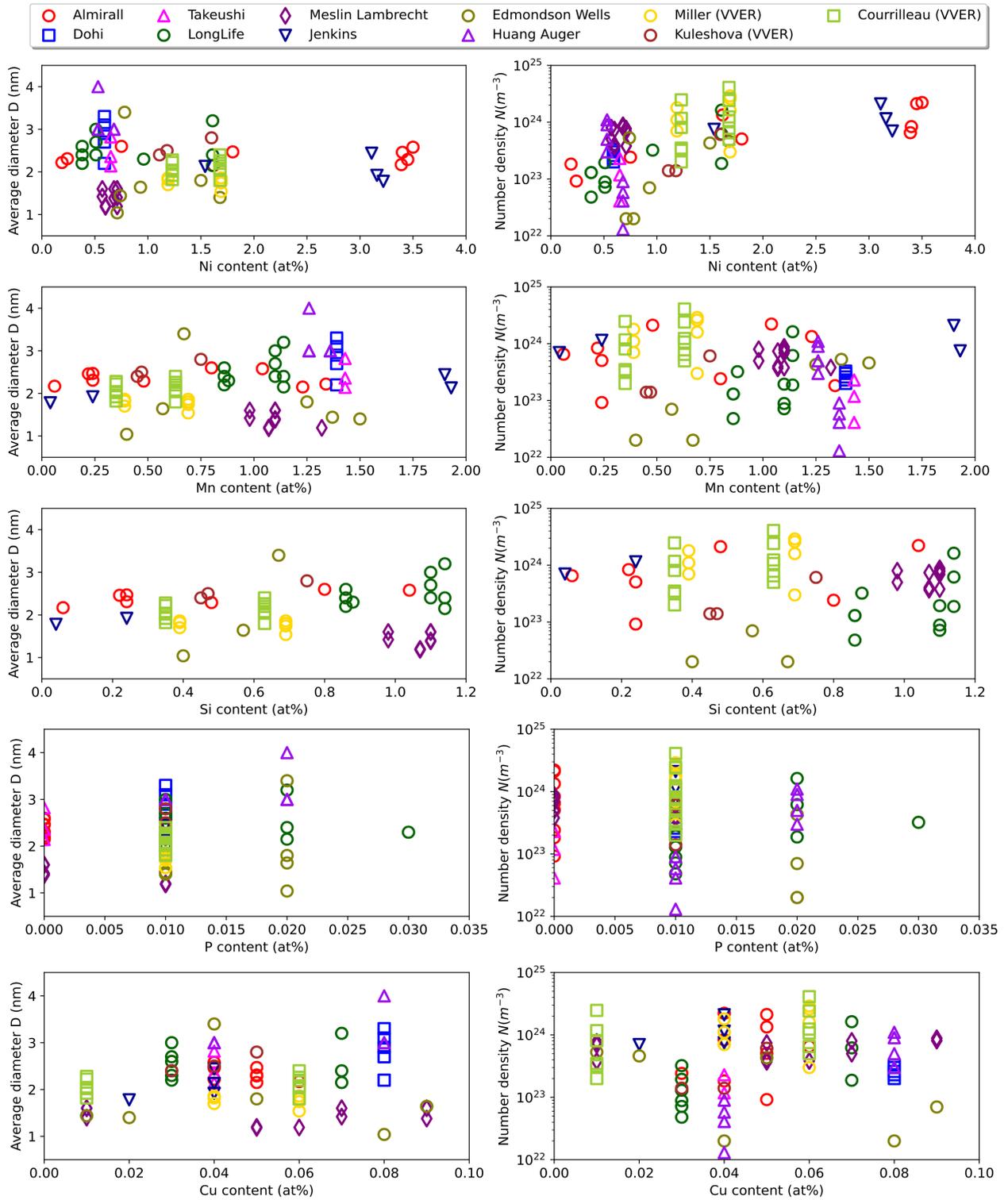

**Fig. SI2 –** (Absence of) correlations between the reported $N$ and $D$ and the chemical contents in the materials included in the various data sets defined in Tab. 1 (main manuscript).

## S2 Additional figures to Fig. 6.

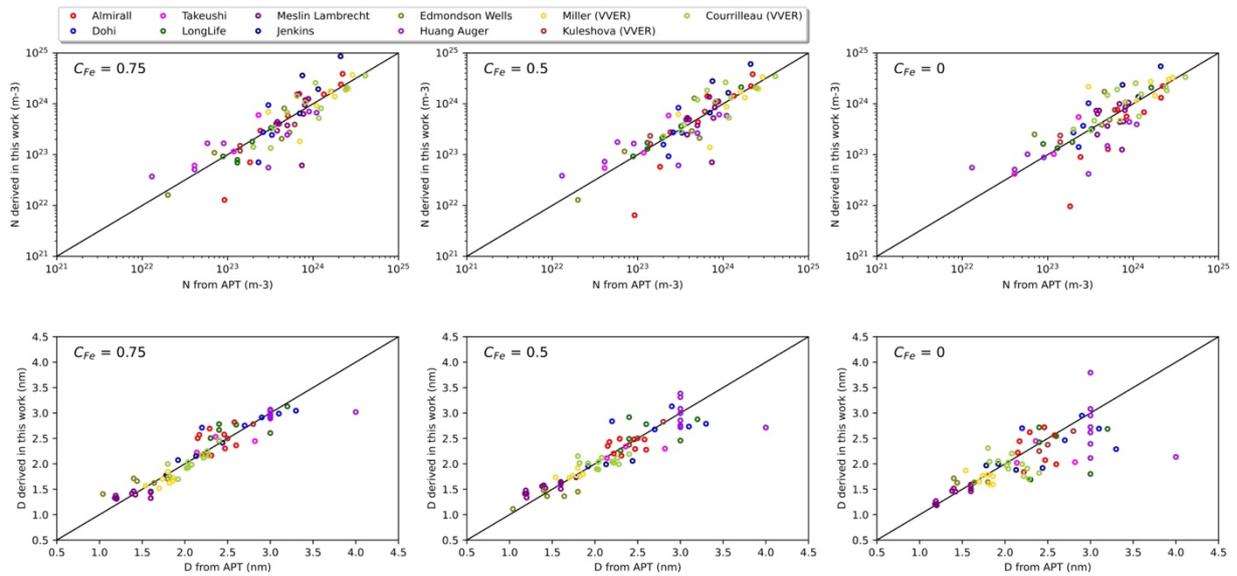

**Fig. SI3** – Reproduction of Fig. 6 (main manuscript), without error bars.

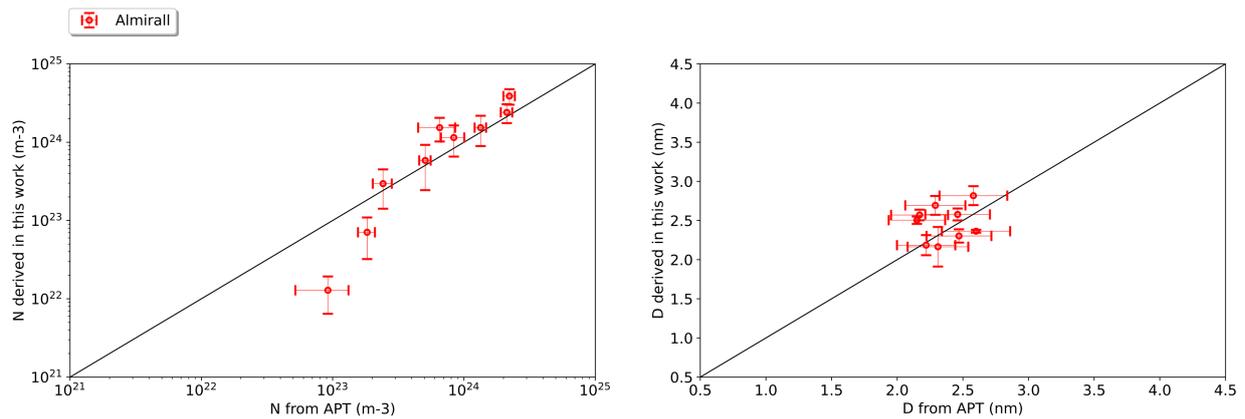

**Fig. SI4** – Reproduction of Fig. 6 (main manuscript), for $C_{Fe}$=0.75, with only the data point from the Almirall set.

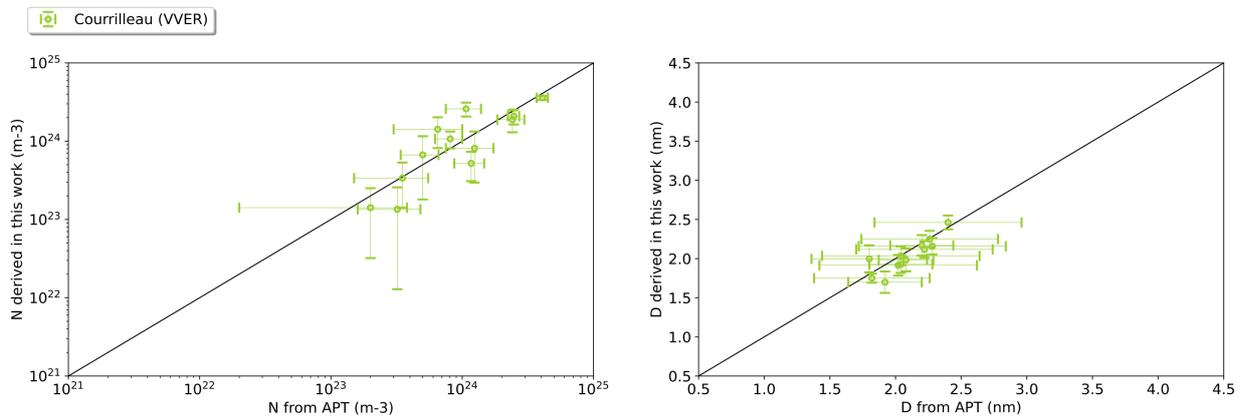

**Fig. SI5** – Reproduction of Fig. 6 (main manuscript), for $C_{Fe}$=0.75, with only the data point from the Courrillea set.

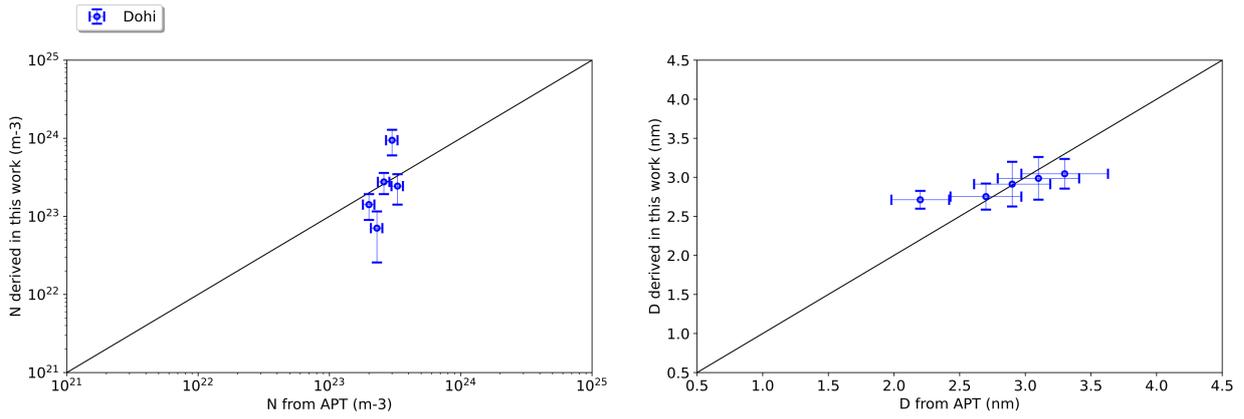

**Fig. SI6 –** Reproduction of Fig. 6 (main manuscript), for $C_{Fe}$=0.75, with only the data point from the Dohi set.

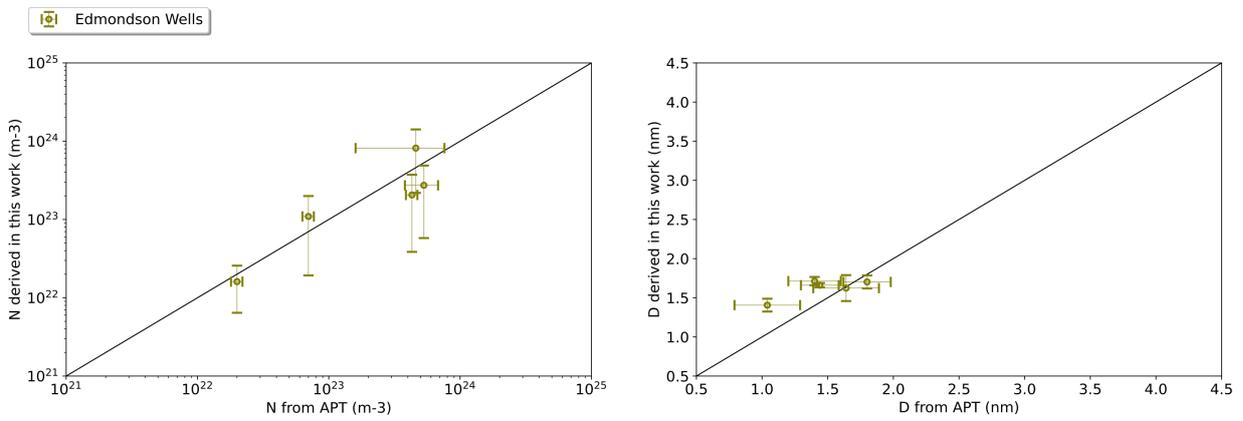

**Fig. SI7 –** Reproduction of Fig. 6 (main manuscript), for $C_{Fe}$=0.75, with only the data point from the Edmondson-Wells set.

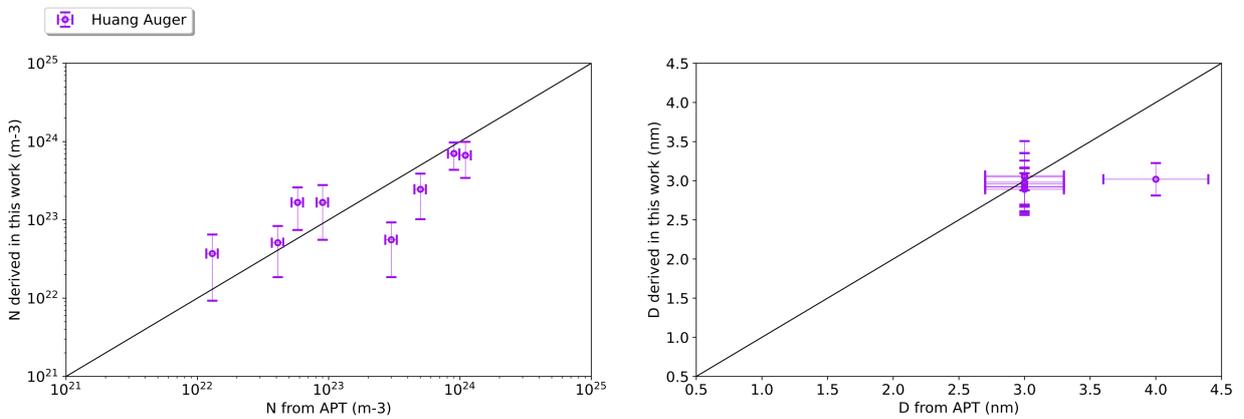

**Fig. SI8 –** Reproduction of Fig. 6 (main manuscript), for $C_{Fe}$=0.75, with only the data point from the Huang-Auger set.

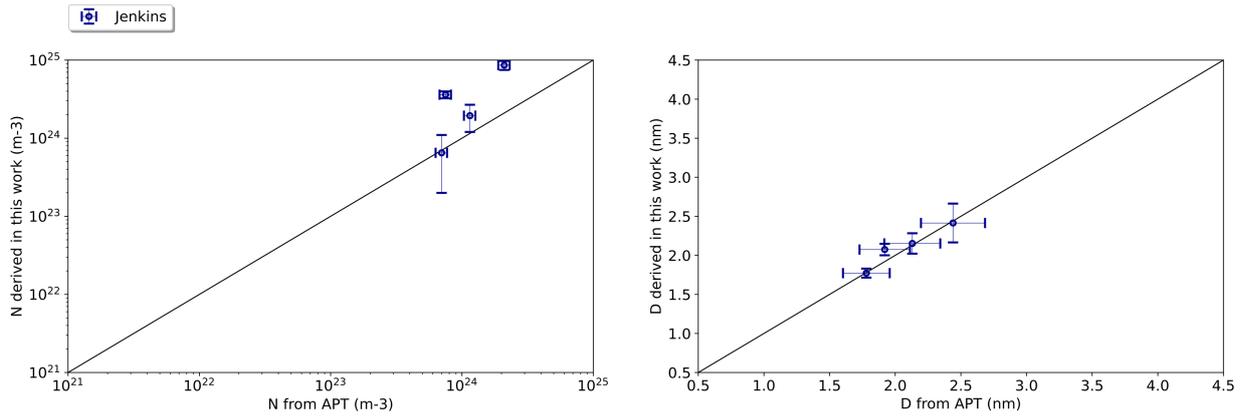

**Fig. SI9 –** Reproduction of Fig. 6 (main manuscript), for $C_{Fe}$=0.75, with only the data point from the Jenkins set.

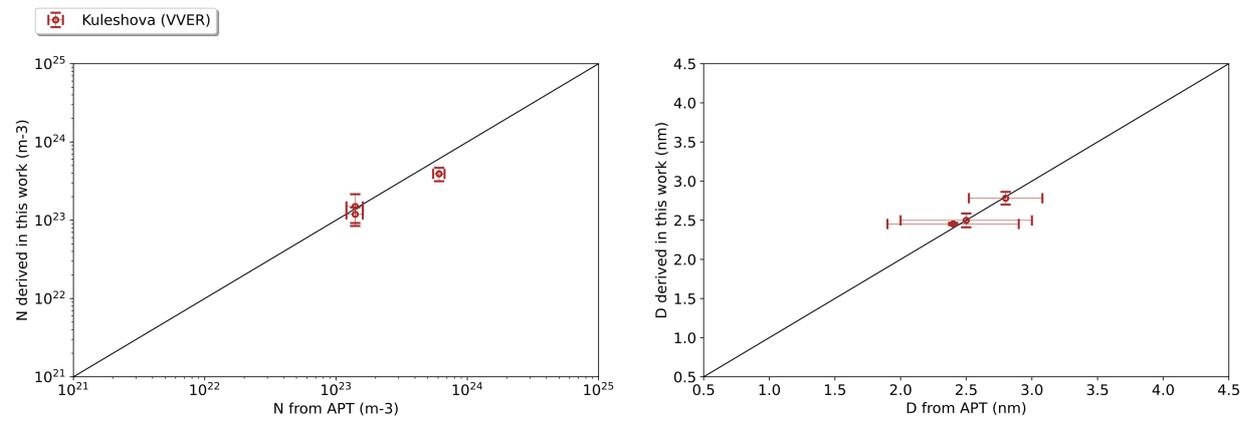

**Fig. SI10 –** Reproduction of Fig. 6 (main manuscript), for $C_{Fe}$=0.75, with only the data point from the Kuleshova set.

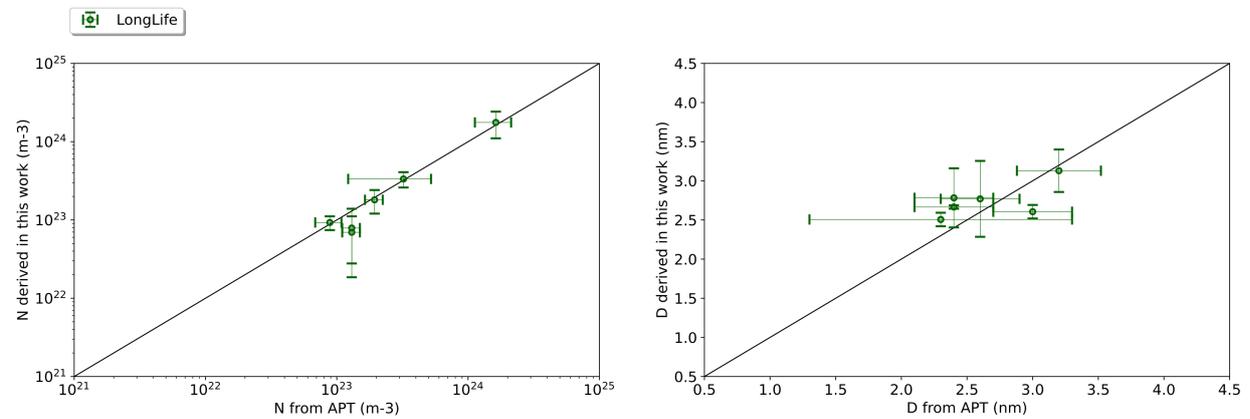

**Fig. SI11 –** Reproduction of Fig. 6 (main manuscript), for $C_{Fe}$=0.75, with only the data point from the LongLife set.

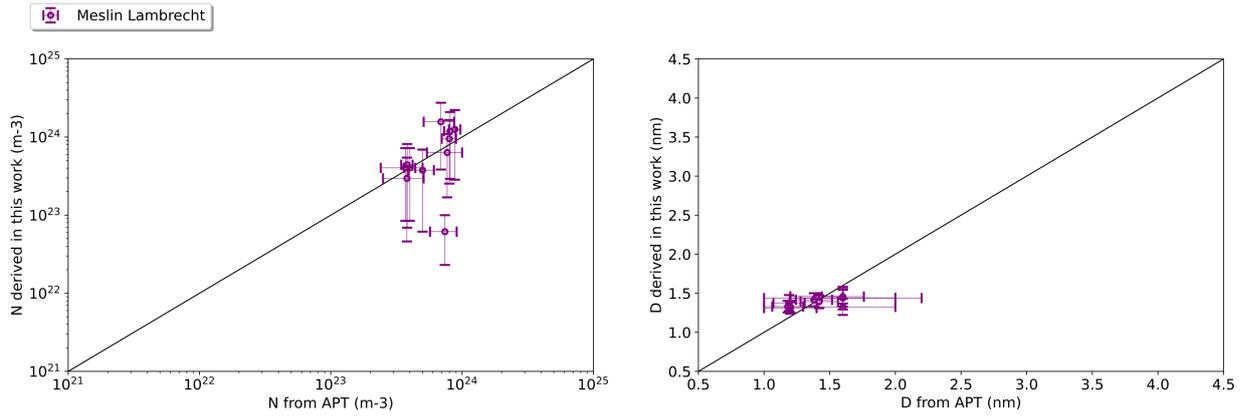

**Fig. SI12 –** Reproduction of Fig. 6 (main manuscript), for $C_{Fe}$=0.75, with only the data point from the Meslin-Lambrecht set.

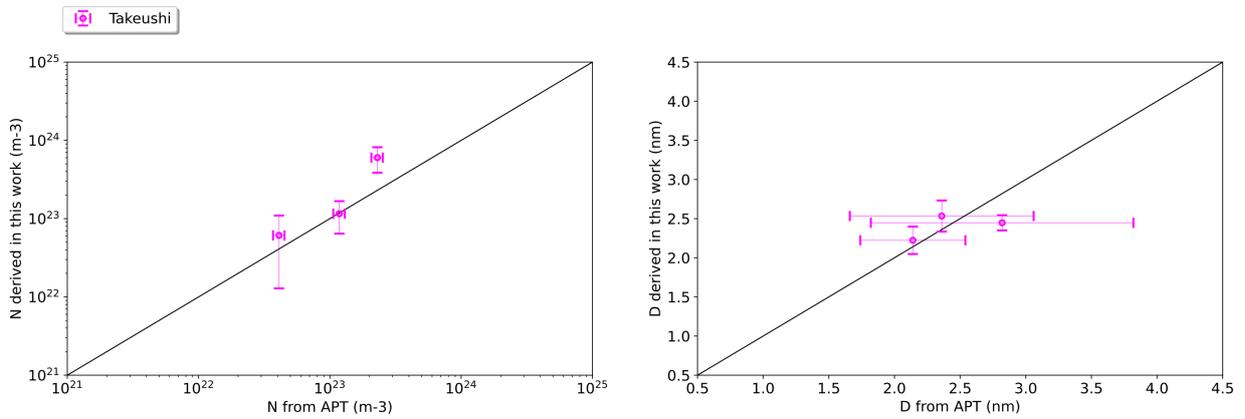

**Fig. SI13 –** Reproduction of Fig. 6 (main manuscript), for $C_{Fe}$=0.75, with only the data point from the Takeushi set.

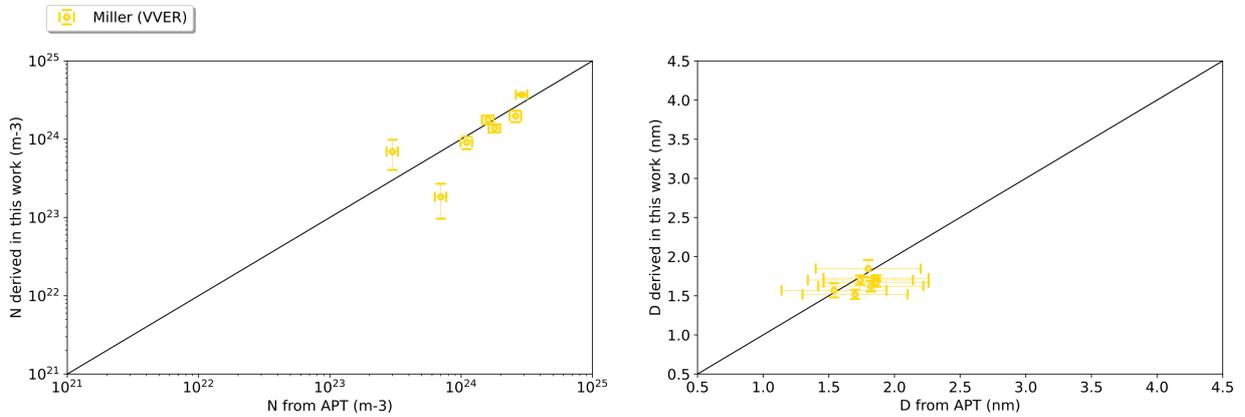

**Fig. SI14 –** Reproduction of Fig. 6 (main manuscript), for $C_{Fe}$=0.75, with only the data point from the Miller set.

**Tab SI1** – Detailed information for the materials and irradiation conditions considered in this work (Part 1/3).

| Set name | Ref | Kind | Name | Materials composition and microstructure (Inputs variables for the OKMC model) | | | | | | | | | Irradiation conditions (Inputs variables for the OKMC model) | | | Experimental APT measurements as reported in the cited references | | | |
|---|---|---|---|---|---|---|---|---|---|---|---|---|---|---|---|---|---|---|---|
| | | | | $C_{Ni}$ (at%) | $C_{Mn}$ (at%) | $C_{Si}$ (at%) | $C_P$ (at%) | $C_{Cu}$ (at%) | $C_{Cr}$ (at%) | $D_{GB}$ (µm) | $C_C$ (appm) | $\rho_d$ (m⁻²) | $T_{irr}$ (°C) | $R_{irr}$ (dpa/s) | Dose (dpa) | N (10²² m-3) | N Error | D (nm) | D Error |
| almirall | [25] | RPV | ALM-R1 | 0.24 | 0.24 | 0.49 | 0 | 0.05 | 0 | 1 | 150 | 2.50E+14 | 290 | 1.00E-08 | 0.2 | 9.2 | 4 | 2.31 | 0.02 |
| almirall | [25] | RPV | ALM-R17 | 3.5 | 1.04 | 0.44 | 0 | 0.04 | 0 | 1 | 150 | 2.50E+14 | 290 | 1.00E-08 | 0.2 | 222 | 1.2 | 2.58 | 0.09 |
| almirall | [25] | RPV | ALM-R19 | 1.8 | 0.24 | 0.47 | 0 | 0.05 | 0 | 1 | 150 | 2.50E+14 | 290 | 1.00E-08 | 0.2 | 50.7 | 0.9 | 2.47 | 0.03 |
| almirall | [25] | RPV | ALM-R22 | 1.62 | 1.23 | 0.46 | 0 | 0.05 | 0 | 1 | 150 | 2.50E+14 | 290 | 1.00E-08 | 0.2 | 134 | 1.5 | 2.15 | 0.08 |
| almirall | [25] | RPV | ALM-R26 | 3.4 | 0.22 | 0.39 | 0 | 0.04 | 0 | 1 | 150 | 2.50E+14 | 290 | 1.00E-08 | 0.2 | 83.7 | 16.9 | 2.46 | 0.04 |
| almirall | [25] | RPV | ALM-R34 | 3.39 | 0.06 | 0.4 | 0 | 0.06 | 0 | 1 | 150 | 2.50E+14 | 290 | 1.00E-08 | 0.2 | 65.3 | 20.5 | 2.17 | 0.1 |
| almirall | [25] | RPV | ALM-R35 | 0.19 | 1.34 | 0.46 | 0 | 0.04 | 0 | 1 | 150 | 2.50E+14 | 290 | 1.00E-08 | 0.2 | 18.3 | 2.7 | 2.22 | 0.03 |
| almirall | [25] | RPV | ALM-R39 | 0.75 | 0.8 | 0.46 | 0 | 0.03 | 0 | 1 | 150 | 2.50E+14 | 290 | 1.00E-08 | 0.2 | 24.2 | 4 | 2.6 | 0.13 |
| almirall | [25] | RPV | ALM-R48 | 3.45 | 0.48 | 0.42 | 0 | 0.05 | 0 | 1 | 150 | 2.50E+14 | 290 | 1.00E-08 | 0.2 | 212 | 6 | 2.29 | 0.08 |
| LongLife | [9] | RPV | ANP2 | 0.96 | 0.88 | 0.24 | 0.03 | 0.03 | 0.11 | 1 | 150 | 2.50E+14 | 285 | 4.00E-09 | 0.09 | 32.21 | 20 | 2.3 | 1 |
| LongLife | [9] | RPV | ANP6 | 1.61 | 1.14 | 0.3 | 0.02 | 0.07 | 0.08 | 1 | 150 | 2.50E+14 | 280 | 4.00E-09 | 0.07 | 162.69 | 50 | 3.2 | 0 |
| LongLife | [9] | RPV | EDF2 | 0.51 | 1.1 | 0.36 | 0.01 | 0.03 | 0.21 | 1 | 150 | 2.50E+14 | 286 | 3.00E-10 | 0.07 | 8.85 | 2 | 2.4 | 0.3 |
| LongLife | [9] | RPV | EDF2 | 0.51 | 1.1 | 0.36 | 0.01 | 0.03 | 0.21 | 1 | 150 | 2.50E+14 | 286 | 3.00E-10 | 0.1 | 19.4 | 3 | 3 | 0.3 |
| LongLife | [9] | RPV | EDF3 | 0.38 | 0.86 | 1 | 0.01 | 0.03 | 0.02 | 1 | 150 | 2.50E+14 | 286 | 3.00E-10 | 0.06 | 13 | 2 | 2.6 | 0.3 |
| LongLife | [9] | RPV | EDF3 | 0.38 | 0.86 | 1 | 0.01 | 0.03 | 0.02 | 1 | 150 | 2.50E+14 | 286 | 3.00E-10 | 0.07 | 13 | 2 | 2.4 | 0.3 |
| Dohi | [40] | RPV | S11 | 0.59 | 1.39 | 0.51 | 0.01 | 0.08 | 0 | 1 | 150 | 2.50E+14 | 290 | 5.44E-10 | 0.05 | 23 | - | 2.9 | - |
| Dohi | [40] | RPV | S11 | 0.59 | 1.39 | 0.51 | 0.01 | 0.08 | 0 | 1 | 150 | 2.50E+14 | 290 | 5.61E-10 | 0.06 | 20 | - | 2.2 | - |
| Dohi | [40] | RPV | S11 | 0.59 | 1.39 | 0.51 | 0.01 | 0.08 | 0 | 1 | 150 | 2.50E+14 | 290 | 1.04E-09 | 0.1 | 33 | - | 2.7 | - |
| Dohi | [40] | RPV | S11 | 0.59 | 1.39 | 0.51 | 0.01 | 0.08 | 0 | 1 | 150 | 2.50E+14 | 290 | 1.11E-09 | 0.11 | 26 | - | 3.1 | - |
| Dohi | [40] | RPV | S11 | 0.59 | 1.39 | 0.51 | 0.01 | 0.08 | 0 | 1 | 150 | 2.50E+14 | 290 | 2.11E-09 | 0.21 | 30 | - | 3.3 | - |
| Takeushi | [41] | RPV | Steel-B | 0.65 | 1.43 | 0.19 | 0 | 0.04 | 0.13 | 1 | 150 | 2.50E+14 | 290 | 2.81E-08 | 0.03 | 4.1 | - | 2.14 | 0.4 |
| Takeushi | [41] | RPV | Steel-B | 0.65 | 1.43 | 0.19 | 0 | 0.04 | 0.13 | 1 | 150 | 2.50E+14 | 290 | 3.16E-08 | 0.07 | 11.8 | - | 2.36 | 0.7 |
| Takeushi | [41] | RPV | Steel-B | 0.65 | 1.43 | 0.19 | 0 | 0.04 | 0.13 | 1 | 150 | 2.50E+14 | 290 | 3.33E-08 | 0.17 | 23 | - | 2.82 | 1 |

**Tab SI2** – Detailed information for the materials and irradiation conditions considered in this work (Part 2/3).

| Set name | Ref | Kind | Name | $C_{Ni}$ (at%) | $C_{Mn}$ (at%) | $C_{Si}$ (at%) | $C_P$ (at%) | $C_{Cu}$ (at%) | $C_{Cr}$ (at%) | $D_{GB}$ (µm) | $C_C$ (appm) | $\rho_d$ (m⁻²) | $T_{irr}$ (°C) | $R_{irr}$ (dpa/s) | Dose (dpa) | N (10²² m-3) | N Error | D (nm) | D Error |
|---|---|---|---|---|---|---|---|---|---|---|---|---|---|---|---|---|---|---|---|
| Edmond.-Wells | [33] | RPV | Forging-Ginna-A | 0.78 | 0.67 | 0.67 | 0.02 | 0.04 | 0.39 | 1 | 150 | 2.50E+14 | 290 | 1.40E-10 | 0.03 | 2 | 0.2 | 3.4 | 1 |
| Edmond.-Wells | [33] | RPV | Forging-Ginna-B | 0.71 | 0.4 | 0.56 | 0.02 | 0.08 | 0.39 | 1 | 150 | 2.50E+14 | 290 | 1.40E-10 | 0.06 | 2 | 0.2 | 1.04 | 0.25 |
| Edmond.-Wells | [33] | RPV | Forging-Ginna-C | 0.93 | 0.57 | 0.66 | 0.02 | 0.09 | 0.39 | 1 | 150 | 2.50E+14 | 290 | 1.40E-10 | 0.1 | 7 | 0.5 | 1.64 | 0.25 |
| Edmond.-Wells | [34] | RPV | Ringhals3 | 1.5 | 1.25 | 0.41 | 0.02 | 0.05 | 0.26 | 1 | 150 | 2.50E+14 | 284 | 2.58E-10 | 0.11 | 43 | 1 | 1.8 | 0.04 |
| Edmond.-Wells | [6] | RPV | CM6 | 1.68 | 1.5 | 0.17 | 0.01 | 0.02 | 0 | 1 | 150 | 2.50E+14 | 300 | 1.50E-09 | 0.2 | 46 | 30 | 1.4 | 0.2 |
| Edmond.-Wells | [6] | RPV | LG | 0.74 | 1.37 | 0.22 | 0.01 | 0.01 | 0 | 1 | 150 | 2.50E+14 | 300 | 1.50E-09 | 0.2 | 53 | 15 | 1.44 | 0.08 |
| Huang-Auger | [26] | RPV | Chooz-A | 0.53 | 1.26 | 0.63 | 0.02 | 0.08 | 0.17 | 1 | 150 | 2.50E+14 | 270 | 2.63E-10 | 0.04 | 30 | - | 3 | - |
| Huang-Auger | [26] | RPV | Chooz-A | 0.53 | 1.26 | 0.63 | 0.02 | 0.08 | 0.17 | 1 | 150 | 2.50E+14 | 270 | 2.63E-10 | 0.12 | 50 | - | 3 | - |
| Huang-Auger | [26] | RPV | Chooz-A | 0.53 | 1.26 | 0.63 | 0.02 | 0.08 | 0.17 | 1 | 150 | 2.50E+14 | 270 | 2.63E-10 | 0.21 | 90 | - | 4 | - |
| Huang-Auger | [26] | RPV | Chooz-A | 0.53 | 1.26 | 0.63 | 0.02 | 0.08 | 0.17 | 1 | 150 | 2.50E+14 | 270 | 2.63E-10 | 0.28 | 110 | - | 3 | - |
| Huang-Auger | [27] | RPV | 16MND5_A | 0.68 | 1.36 | 0.53 | 0.01 | 0.04 | 0.24 | 1 | 150 | 2.50E+14 | 288 | 3.51E-10 | 0.03 | 1.3 | - | 3 | - |
| Huang-Auger | [27] | RPV | 16MND5_A | 0.68 | 1.36 | 0.53 | 0.01 | 0.04 | 0.24 | 1 | 150 | 2.50E+14 | 288 | 3.51E-10 | 0.06 | 4.1 | - | 3 | - |
| Huang-Auger | [27] | RPV | 16MND5_A | 0.68 | 1.36 | 0.53 | 0.01 | 0.04 | 0.24 | 1 | 150 | 2.50E+14 | 288 | 3.51E-10 | 0.1 | 5.8 | - | 3 | - |
| Huang-Auger | [27] | RPV | 16MND5_A | 0.68 | 1.36 | 0.53 | 0.01 | 0.04 | 0.24 | 1 | 150 | 2.50E+14 | 288 | 3.51E-10 | 0.13 | 9 | - | 3 | - |
| Jenkins | [28] | RPV | C-A32 | 3.11 | 1.9 | 0.37 | 0.01 | 0.04 | 0.1 | 1 | 150 | 2.50E+14 | 290 | 4.37E-09 | 0.17 | 210 | 1 | 2.44 | 0.19 |
| Jenkins | [28] | RPV | K-A26 | 3.16 | 0.24 | 0.38 | 0.01 | 0.04 | 0.1 | 1 | 150 | 2.50E+14 | 290 | 4.37E-09 | 0.17 | 115 | 1.5 | 1.92 | 0.03 |
| Jenkins | [28] | RPV | RX12-AX12 | 3.22 | 0.04 | 0.06 | 0.01 | 0.02 | 1.84 | 1 | 150 | 2.50E+14 | 290 | 4.37E-09 | 0.17 | 70 | 4.2 | 1.78 | 0.14 |
| Jenkins | [28] | RPV | G-A31 | 1.54 | 1.93 | 0.38 | 0.01 | 0.04 | 0.1 | 1 | 150 | 2.50E+14 | 290 | 4.37E-09 | 0.17 | 75 | 0.4 | 2.13 | 0.09 |
| Meslin- Lambr. | [31] | Model | FeMnNi_B | 0.71 | 1.1 | 0.01 | 0 | 0.01 | 0 | 50 | 50 | 1.00E+14 | 290 | 2.50E-09 | 0.03 | 38 | 13 | 1.6 | 0.4 |
| Meslin- Lambr. | [31] | Model | FeMnNi_B | 0.71 | 1.1 | 0.01 | 0 | 0.01 | 0 | 50 | 50 | 1.00E+14 | 290 | 1.29E-07 | 0.1 | 77 | 23 | 1.4 | 0.2 |
| Meslin- Lambr. | [31] | Model | FeMnNi_B | 0.71 | 1.1 | 0.01 | 0 | 0.01 | 0 | 50 | 50 | 1.00E+14 | 290 | 1.43E-07 | 0.2 | 69 | 18 | 1.6 | 0.6 |
| Meslin- Lambr. | [29] | Model | FeMnNiCu | 0.68 | 1.1 | 0.01 | 0 | 0.09 | 0 | 50 | 50 | 1.00E+14 | 290 | 1.55E-07 | 0.1 | 88.2 | - | 1.38 | - |
| Meslin- Lambr. | [31] | Model | FeMnNi_A | 0.68 | 1.1 | 0.01 | 0 | 0.09 | 0 | 50 | 50 | 1.00E+14 | 290 | 1.55E-07 | 0.1 | 81.11 | - | 1.6 | - |
| Meslin- Lambr. | [29] | RPV | 16MND5_B | 0.71 | 1.32 | 0.39 | 0.01 | 0.06 | 0 | 1 | 150 | 2.50E+14 | 300 | 1.55E-07 | 0.1 | 38.18 | - | 1.19 | - |
| Meslin- Lambr. | [30,31] | RPV | 16MND5_C | 0.6 | 1.07 | 0.48 | 0.01 | 0.05 | 0 | 1 | 150 | 2.50E+14 | 300 | 1.40E-07 | 0.2 | 40 | 0 | 1.18 | 0.1 |
| Meslin- Lambr. | [31,32] | Model | FeCuMnNi | 0.57 | 0.98 | 0 | 0 | 0.07 | 0 | 50 | 50 | 1.00E+14 | 290 | 1.40E-07 | 0.05 | 50 | 11 | 1.42 | 0.1 |
| Meslin- Lambr. | [31,32] | Model | FeCuMnNi | 0.57 | 0.98 | 0 | 0 | 0.07 | 0 | 50 | 50 | 1.00E+14 | 290 | 1.40E-07 | 0.1 | 80 | 10 | 1.6 | 0.4 |
| Meslin- Lambr. | [30,31] | RPV | 16MND5_C | 0.6 | 1.07 | 0.48 | 0.01 | 0.05 | 0 | 1 | 150 | 2.50E+14 | 300 | 2.40E-09 | 0.03 | 74 | 17 | 1.2 | 0.2 |
| Meslin- Lambr. | [30,31] | RPV | 16MND5_C | 0.6 | 1.07 | 0.48 | 0.01 | 0.05 | 0 | 1 | 150 | 2.50E+14 | 300 | 1.40E-07 | 0.2 | 37 | 13 | 1.2 | 0.2 |

**Tab SI3** – Detailed information for the materials and irradiation conditions considered in this work (Part 3/3).

| Set name | Ref | Kind | Name | Materials composition and microstructure (Inputs variables for the OKMC model) | | | | | | | | | Irradiation conditions (Inputs variables for the OKMC model) | | | Experimental APT measurements as reported in the cited references | | | |
|---|---|---|---|---|---|---|---|---|---|---|---|---|---|---|---|---|---|---|---|
| | | | | $C_{Ni}$ (at%) | $C_{Mn}$ (at%) | $C_{Si}$ (at%) | $C_P$ (at%) | $C_{Cu}$ (at%) | $C_{Cr}$ (at%) | $D_{GB}$ (μm) | $C_C$ (appm) | $\rho_d$ (m$^{-2}$) | $T_{irr}$ (°C) | $R_{irr}$ (dpa/s) | Dose (dpa) | N (10$^{22}$ m-3) | N Error | D (nm) | D Error |
| Kuleshova | [37] | VVER | 1000BM_A | 1.18 | 0.47 | 0.6 | 0.01 | 0.04 | 2.34 | 10 | 150 | 2.50E+14 | 300 | 8.23E-08 | 0.08 | 14 | 2 | 2.5 | 0.5 |
| Kuleshova | [38] | VVER | 1000W-RS3 | 1.6 | 0.75 | 0.28 | 0.01 | 0.05 | 2 | 10 | 150 | 2.50E+14 | 300 | 2.19E-10 | 0.08 | 61 | - | 2.8 | - |
| Kuleshova | [39] | VVER | 15Kh2NMFAA | 1.11 | 0.45 | 0.57 | 0.01 | 0.03 | 2.22 | 10 | 150 | 2.50E+14 | 300 | 1.16E-08 | 0.08 | 14 | 2 | 2.4 | 0.5 |
| Miller | [36] | VVER | 1000BM_B | 1.19 | 0.39 | 0.59 | 0.01 | 0.04 | 2.34 | 10 | 150 | 2.50E+14 | 290 | 8.77E-10 | 0.04 | 70 | 7 | 1.7 | 0.4 |
| Miller | [36] | VVER | 1000BM_B | 1.19 | 0.39 | 0.59 | 0.01 | 0.04 | 2.34 | 10 | 150 | 2.50E+14 | 290 | 5.26E-09 | 0.17 | 110 | 11 | 1.82 | 0.4 |
| Miller | [36] | VVER | 1000BM_B | 1.19 | 0.39 | 0.59 | 0.01 | 0.04 | 2.34 | 10 | 150 | 2.50E+14 | 290 | 5.26E-09 | 0.26 | 180 | 18 | 1.86 | 0.4 |
| Miller | [36] | VVER | 1000W | 1.69 | 0.69 | 0.65 | 0.01 | 0.06 | 1.93 | 10 | 150 | 2.50E+14 | 290 | 8.77E-10 | 0.04 | 30 | 3 | 1.54 | 0.4 |
| Miller | [36] | VVER | 1000W | 1.69 | 0.69 | 0.65 | 0.01 | 0.06 | 1.93 | 10 | 150 | 2.50E+14 | 290 | 5.26E-09 | 0.09 | 160 | 16 | 1.74 | 0.4 |
| Miller | [36] | VVER | 1000W | 1.69 | 0.69 | 0.65 | 0.01 | 0.06 | 1.93 | 10 | 150 | 2.50E+14 | 290 | 5.26E-09 | 0.12 | 260 | 26 | 1.86 | 0.4 |
| Miller | [36] | VVER | 1000W | 1.69 | 0.69 | 0.65 | 0.01 | 0.06 | 1.93 | 10 | 150 | 2.50E+14 | 290 | 5.26E-09 | 0.2 | 290 | 29 | 1.8 | 0.4 |
| Courrilleau | [35] | VVER | 1000BM_LF265 | 1.23 | 0.35 | 0.63 | 0.01 | 0.01 | 2.5 | 10 | 150 | 2.50E+14 | 265 | 1.50E-08 | 0.03 | 32 | 16 | 1.92 | 0.28 |
| Courrilleau | [35] | VVER | 1000BM_MF265 | 1.23 | 0.35 | 0.63 | 0.01 | 0.01 | 2.5 | 10 | 150 | 2.50E+14 | 265 | 4.50E-08 | 0.1 | 117 | 30 | 2.04 | 0.24 |
| Courrilleau | [35] | VVER | 1000BM_HF265 | 1.23 | 0.35 | 0.63 | 0.01 | 0.01 | 2.5 | 10 | 150 | 2.50E+14 | 265 | 1.15E-07 | 0.27 | 247 | 24 | 2.28 | 0.56 |
| Courrilleau | [35] | VVER | 1000BM_LF300 | 1.23 | 0.35 | 0.63 | 0.01 | 0.01 | 2.5 | 10 | 150 | 2.50E+14 | 300 | 1.65E-08 | 0.03 | 20 | 18 | 1.82 | 0.44 |
| Courrilleau | [35] | VVER | 1000BM_MF300 | 1.23 | 0.35 | 0.63 | 0.01 | 0.01 | 2.5 | 10 | 150 | 2.50E+14 | 300 | 4.95E-08 | 0.08 | 35 | 20 | 2.02 | 0.6 |
| Courrilleau | [35] | VVER | 1000BM_HF300 | 1.23 | 0.35 | 0.63 | 0.01 | 0.01 | 2.5 | 10 | 150 | 2.50E+14 | 300 | 1.26E-07 | 0.21 | 81 | 19 | 2.22 | 0.52 |
| Courrilleau | [35] | VVER | 1000W_LF265 | 1.68 | 0.63 | 0.5 | 0.01 | 0.06 | 2.5 | 10 | 150 | 2.50E+14 | 265 | 2.10E-08 | 0.05 | 124 | 49 | 2.08 | 0.08 |
| Courrilleau | [35] | VVER | 1000W_MF265 | 1.68 | 0.63 | 0.5 | 0.01 | 0.06 | 2.5 | 10 | 150 | 2.50E+14 | 265 | 5.40E-08 | 0.12 | 241 | 56 | 2.2 | 0.24 |
| Courrilleau | [35] | VVER | 1000W_HF265 | 1.68 | 0.63 | 0.5 | 0.01 | 0.06 | 2.5 | 10 | 150 | 2.50E+14 | 265 | 1.21E-07 | 0.28 | 409 | 23 | 2.4 | 0.56 |
| Courrilleau | [35] | VVER | 1000W_LF300 | 1.68 | 0.63 | 0.5 | 0.01 | 0.06 | 2.5 | 10 | 150 | 2.50E+14 | 300 | 2.25E-08 | 0.04 | 50 | 16 | 1.8 | 0.44 |
| Courrilleau | [35] | VVER | 1000W_MF300 | 1.68 | 0.63 | 0.5 | 0.01 | 0.06 | 2.5 | 10 | 150 | 2.50E+14 | 300 | 5.85E-08 | 0.1 | 65 | 35 | 2.04 | 0.6 |
| Courrilleau | [35] | VVER | 1000W_HF300 | 1.68 | 0.63 | 0.5 | 0.01 | 0.06 | 2.5 | 10 | 150 | 2.50E+14 | 300 | 1.32E-07 | 0.22 | 107 | 32 | 2.26 | 0.52 |